\documentclass[a4paper,twocolumn,11pt,allowtoday,accepted=2021-03-06]{quantumarticle}
\pdfoutput=1
\usepackage[numbers,sort&compress]{natbib}
\usepackage{textcomp}
\usepackage{amsmath}
\usepackage{amssymb}
\usepackage{graphicx}
\usepackage{enumitem}
\usepackage{bm}
\usepackage{braket}
\usepackage[breaklinks]{hyperref}
\hypersetup{colorlinks=true, linkcolor=blue, citecolor=blue, filecolor=blue, urlcolor=blue}
\usepackage{amsthm}
\usepackage{enumitem}
\usepackage{microtype}

\begin{document}

\title{Hyper-optimized tensor network contraction}

\author{Johnnie Gray}

\affiliation{Blackett Laboratory, Imperial College London, London SW7 2AZ, United Kingdom}
\affiliation{Division of Chemistry and Chemical Engineering, California Institute of Technology, Pasadena, California 91125, USA}

\author{Stefanos Kourtis}

\affiliation{Blackett Laboratory, Imperial College London, London SW7 2AZ, United Kingdom}
\affiliation{Department of Physics, Boston University, Boston, MA, 02215, USA}
\affiliation{Institut quantique \& D\'{e}partement de physique, Universit\'{e} de Sherbrooke, Qu\'{e}bec J1K 2R1, Canada}

\date{\rm\today}

\begin{abstract}
Tensor networks represent the state-of-the-art in computational methods across many disciplines, including the classical simulation of quantum many-body systems and quantum circuits. Several applications of current interest give rise to tensor networks with irregular geometries. Finding the best possible contraction path for such networks is a central problem, with an exponential effect on computation time and memory footprint. In this work, we implement new randomized protocols that find very high quality contraction paths for arbitrary and large tensor networks. We test our methods on a variety of benchmarks, including the random quantum circuit instances recently implemented on Google quantum chips. We find that the paths obtained can be very close to optimal, and often many orders or magnitude better than the most established approaches. As different underlying geometries suit different methods, we also introduce a hyper-optimization approach, where both the method applied and its algorithmic parameters are tuned during the path finding. The increase in quality of contraction schemes found has significant practical implications for the simulation of quantum many-body systems and particularly for the benchmarking of new quantum chips.
Concretely, we estimate a speed-up of over 10,000$\times$ compared to the original expectation for the classical simulation of the Sycamore `supremacy' circuits.
\end{abstract}

\maketitle

\section{Introduction}\label{sec:intro}

Since the advent of the density-matrix renormalization group algorithm, invented to study one-dimensional lattice systems of quantum degrees of freedom, \emph{tensor networks} have permeated a plethora of scientific disciplines, finding use in fields such as quantum condensed matter~\cite{Verstraete2008a, Orus2014, Bridgeman2017, Biamonte2017}, classical statistical mechanics~\cite{Levin2007a, Evenbly2015, Evenbly2017}, information science and big-data processing~\cite{Cichocki2016, Cichocki2017a}, systems engineering~\cite{Duenas-Osorio2018}, quantum computation~\cite{Markov2008}, machine learning and artificial reasoning~\cite{Stoudenmire2016, Stoudenmire2018, Roberts2019} and more. The underlying idea of tensor network methods is to use sparse networks of interconnected low-rank tensors to represent data structures that would otherwise be expressed in (very) high-rank tensor form, which is hard to manipulate. Due to this ubiquity, techniques to perform (multi)linear algebraic operations on tensor networks accurately and efficiently are very useful to a highly interdisciplinary community of researchers and engineers. Of these operations, \emph{tensor network contraction}, i.e., the evaluation of a scalar quantity that has been expressed as a tensor network, is the most common.

When a system under consideration gives rise to a tensor networks with a regular structure, such as lattices, the renormalization group apparatus is often employed to perform tensor network contractions with controllable accuracy. This approach has been successful in tackling a variety of classical and quantum many-body problems~\cite{Levin2007a, Jiang2008, Gu2009, Xie2012, Evenbly2015, Evenbly2017, Zhao2016, Bal2017, Yang2017a}. Efficient tensor network contraction is also possible in special cases in which network topology (e.g., trees), values of tensor entries, or both are restricted~\cite{Shi2006, Valiant2008, Bravyi2008, Aguado2008, Konig2009, Denny2012}. Despite these results, contracting tensor networks with arbitrary structure remains (at least) \#P-hard in the general case~\cite{Valiant1979a,Damm2002}. This is true, in particular, for tensor networks that model \emph{random quantum circuits}, a fact that has recently inspired proposals for quantum algorithms running on these circuits that aim towards a practically demonstrable \emph{quantum computational advantage} over classical computers~\cite{Markov2008, Terhal2004, Bremner2010, Aaronson2013a, Jozsa2013a, Morimae2014, Carolan2015, Farhi2016, Aaronson2016a, Aaronson2016, Boixo2018, Bouland2019}. The key idea is that, unlike quantum algorithms (e.g., Shor or Grover) that require deep quantum circuits and high gate fidelities --- inaccessible in the near future --- to become manifestly advantageous, the task of sampling bit strings from the output of random quantum circuits is expected to be hard to simulate classically even for low-depth circuits and low-fidelity gates. The precise threshold for observing such a quantum advantage is nonuniversal and ultimately depends on the efficiency of the classical simulation for each particular combination of circuit model and quantum chip architecture. This motivates the development of high-performance simulation techniques for these quantum systems, predominantly based on finding good \emph{contraction paths} for tensor networks, that runs in parallel to the race for the development of higher qubit count and quality devices~\cite{Fried2018, Chen2018, Dumitrescu2018}.

\begin{figure}[tb]
    \centering
    \includegraphics[width=\linewidth]{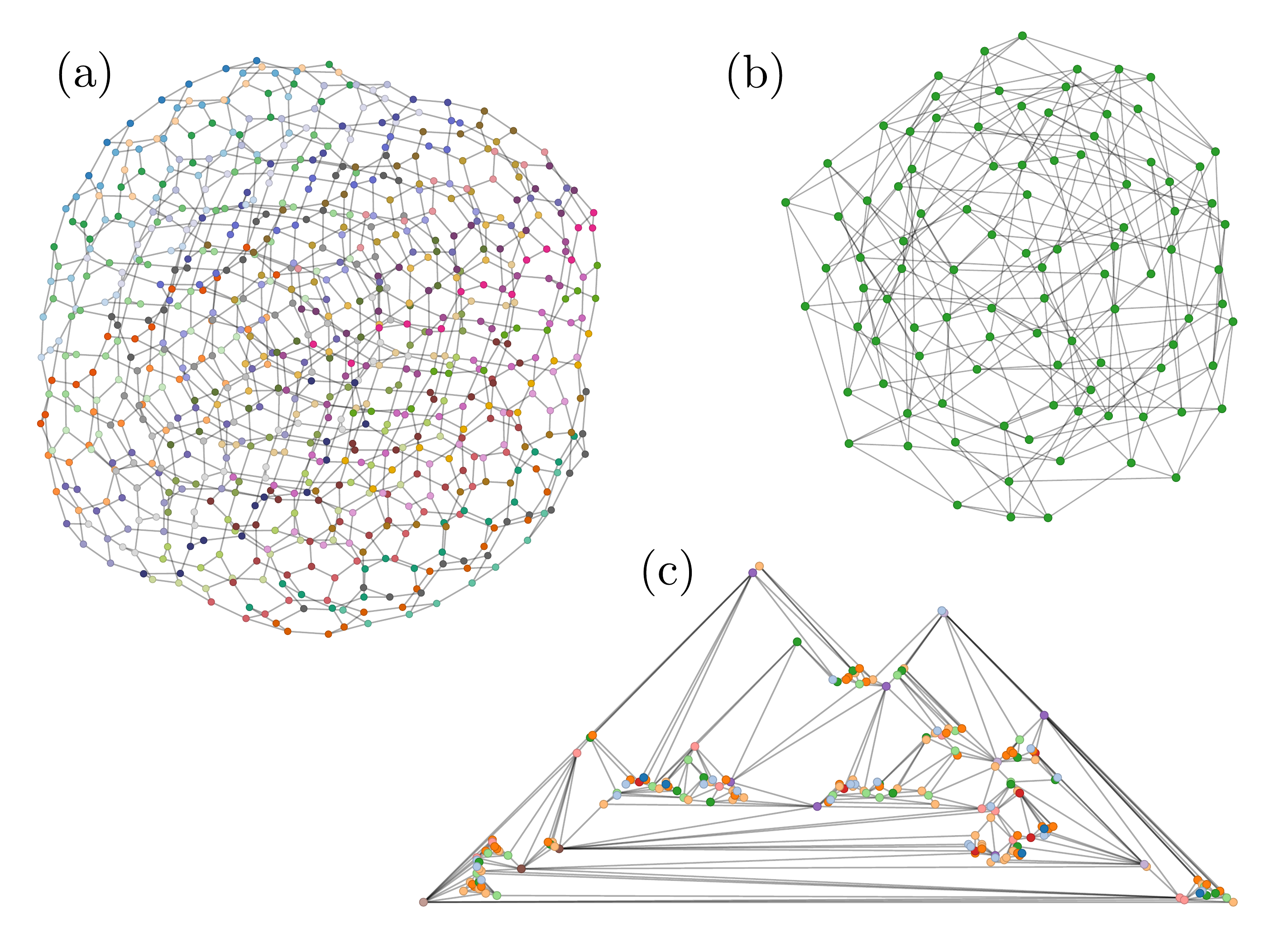}
    \caption{
        Sample tensor networks: (a) simplified network for a rectangular 7x7 qubit 1 + 40 + 1 depth random quantum circuit with 742 rank-3 tensors; (b) a random 5-regular network with 100 tensors, arising in, e.g., SAT problems; and (c) random planar network with 184 tensors, arising in, e.g., the statistical-mechanical evaluation of knot invariants.
    }
    \label{fig:pic-graphs}
\end{figure}

Inspired by the classical simulation of quantum circuits, here we introduce a new framework for exact contraction of large tensor networks with arbitrary structure (see examples in Fig.~\ref{fig:pic-graphs}). The first key idea of this framework is to explicitly construct the \emph{contraction tree} for a given tensor network, combining agglomerative, divisive, and optimal drivers for forming sub-trees at different scales. The second key idea is to hyper-optimize the generation of these trees, and to do this with respect to the entire tree and thus the total contraction cost, rather than just the leading scaling, given by the line-graph tree-width for example. We also establish a powerful set of simplifications for efficiently pre-processing tensor networks prior to contraction.

Using this framework we are able to find very high-quality contraction paths, achieving speedups that scale exponentially with the number of tensors in the network compared to established approaches, for a variety of problems. The drivers we test include recently introduced contraction algorithms based on graph partitioning and community structure detection~\cite{Kourtis2018}, previously theorized~\cite{Markov2008} and recently implemented~\cite{Dudek2019} algorithms based on the tree decomposition of graphs, as well as new heuristics that we introduce in this work. Furthermore, observing that different graph structures favor different algorithms, we implement a hyper-optimization approach, where both the method applied and its parameters are varied throughout the contraction path search, leading to automatically customized contraction algorithms that often achieve near-optimal performance.

We demonstrate the new methodology introduced here on a range of benchmarks. First, we test on problems defined on random graph families, such as simulation of solving MAX-CUT with quantum approximate optimization as well as weighted model counting. We find substantial improvements in performance compared to previous methods reported in the literature. We then simulate random quantum circuits recently implemented by Google on the Bristlecone and Sycamore architectures. We estimate a speed-up of over 10,000$\times$ in the classical simulation of the Sycamore `supremacy' circuits compared to what is given in~\cite{Arute2019}. In general, our algorithms outperform all others for the same task, by a wide margin on general networks and by a narrower margin on planar structures. These findings thus illustrate that our methods can lead to significant performance gains across a spectrum of tensor network applications. This is the main result of this paper.

The remainder of this paper is organized as follows. In Sec.~\ref{sec:problem} we formalize the problem of finding the optimal contraction path for arbitrary tensor networks. In Sec.~\ref{sec:algos} we introduce and explain the various algorithms employed in our heuristics. In Sec.~\ref{sec:results} we test our methods on a variety of benchmarks, including the random quantum circuit instances recently implemented on Google Bristlecone and Sycamore quantum chips, the simulation of the quantum adiabatic optimization algorithm for solving the MAX-CUT problem on random regular graphs, and exact weighted model counting on problem instances from a recent competition. We conclude in Sec.~\ref{sec:conclusion}.

\section{Problem statement}\label{sec:problem}


We denote an \emph{edge-weighted graph} by $G=(V,E)$, where $V$ is the vertex set and the set of 2-tuples of vertex indices $E \subset \{ ( u,v ) : u,v \in V \}$ is the edge set, along with a weight function $w : E\to\mathbb{R}^+$ that assigns a positive real number to each edge. For each vertex $v$, define the \emph{incidence set} $s_v = \{ e : e \in E \ \mathrm{and} \ v \in e \}$, which is the set of edges incident to vertex $v$, such that $|s_v|=d_v$, the \emph{degree} of vertex $v$.

To define a tensor network, we augment $G$ with (i) a discrete variable $x_e$ for each edge $e\in E$, whose set of possible values is given by $D(e)$ with $|D(e)|=w(e)$, (ii) an ordered tuple $t_v : \mathbb{N}_{d_v} \to s_v$ for each vertex $v\in V$, and (iii) a multivariate function or \emph{tensor} $T_v : D(t_v(1)) \times\dots\times D(t_v(d_v)) \to \mathbb{C}$, where $t_v(i)$ denotes the $i$th element of tuple $t_v$, for every vertex $v \in V$. That $w$ is defined to be a real-valued function even though $D(e)\in\mathbb{Z}^+ \ \forall \ e\in E$ is simply a choice that allows for extra flexibility in the design of contraction algorithms, see, e.g., the Boltzmann greedy algorithm below.

With these definitions, a \emph{tensor network contraction} can be represented as a \emph{sequence of vertex contractions} in graph $G$. Each vertex contraction removes common edges between pairs of tensors, if any, and represents a product operation on the corresponding tensors, in which one takes the inner product over common indices or an outer product if there are no common indices. For simplicity, in what follows we consider only \emph{pairwise} contractions, which are common practice. Multiway contractions are also possible, but they can always be decomposed to sequences of pairwise contractions. For some applications, only a subset of $V$ must be contracted, while in others all vertices in $V$ are contracted into a single vertex. Here we will focus on the latter case, as it underlies the former. We will assume that $G$ initially has no loops, i.e., edges connecting vertices to themselves, and that multiple edges are always contracted simultaneously, so that no loops occur throughout the contraction.

To represent the sequence of vertex contractions, we define a \emph{rooted binary tree} $B = (V_B,E_B)$, with the first $|V|$ vertex indices denoting leaves, using two tuples $l$ and $r$ such that $l(v)$ and $r(v)$ are the indices of the `left' and `right' children of vertex $v \in V_B$, respectively, if any. This defines a tree embedding of $G$~\cite{Bienstock1990}. Finally, we assign an incidence set $s_v$ to each $v \in V_B$, starting with leaves, according to
\begin{equation}
s_v = \begin{cases} \{e : e \in E \ \mathrm{and} \ v \in e\} \ \mathrm{if} \ v \ \mathrm{is\ a\ leaf\ index} \,,\\
                    s_{l(v)} \oplus s_{r(v)} \ \mathrm{otherwise} \,, \end{cases}\label{eq:incidence}
\end{equation}
with $s_i \oplus s_j = (s_i \cup s_j) \setminus (s_i \cap s_j)$. The composite of $(B,S)$, where $S = \{s_v : v \in V_B \}$, defines a \emph{contraction tree} of $G$.

\begin{figure}[tb]
    \centering
    \includegraphics[width=\linewidth]{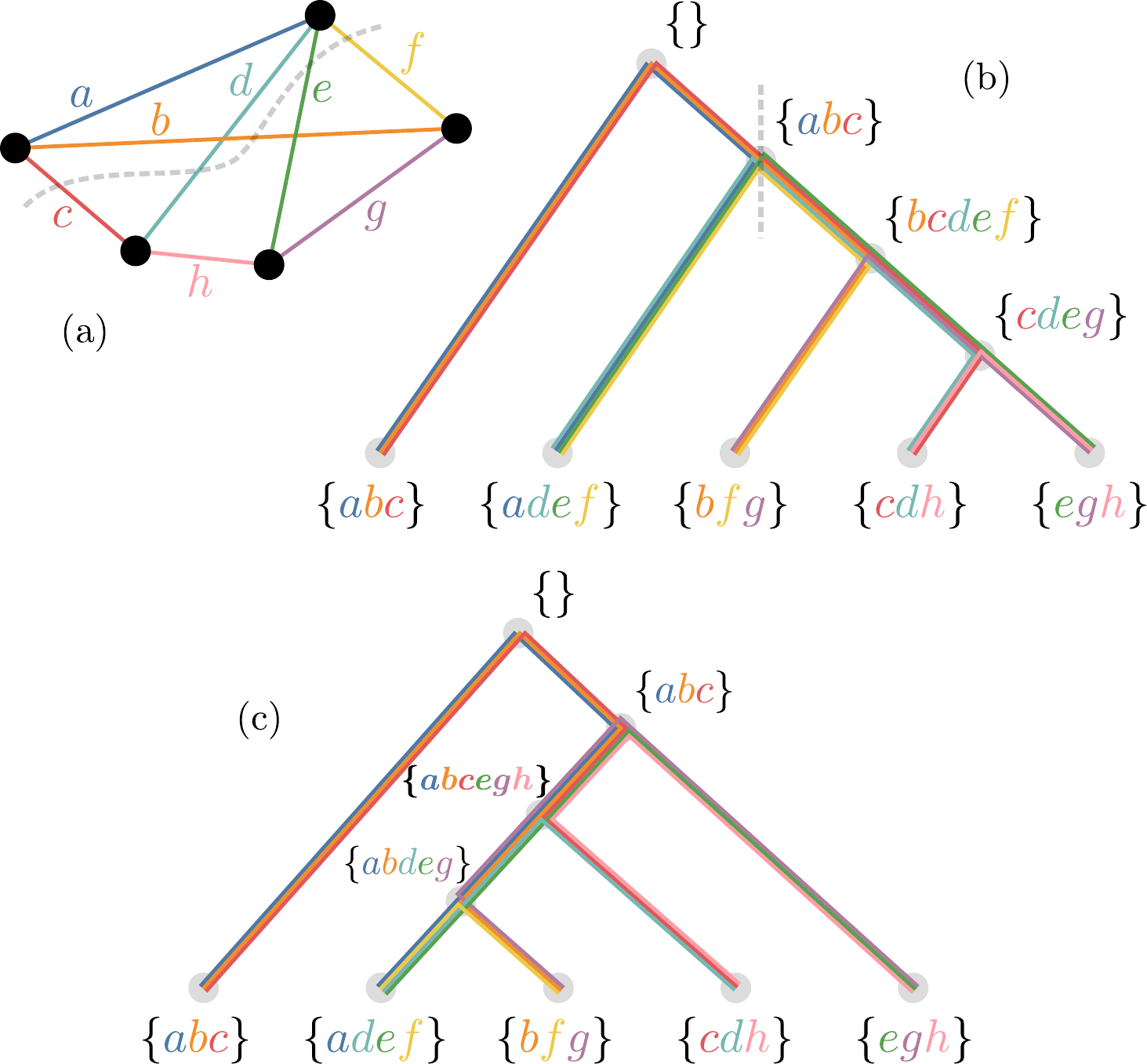}
    \caption{For the graph shown in (a), two possible contraction trees (b) and (c), showing intermediate tensors and congestions.
    Each edge in a tree has an associated tensor and subgraph.
    The size of the tensor is exponential in the number of indices (denoted by unique colors) running along that edge --- the \emph{edge congestion}.
    Each vertex in a tree represents a pairwise contraction of two tensors, as well as a bi-partitioning of the parent edge's subgraph (the dashed grey line shows one example of this).
    The cost of that pairwise contraction is exponential in the number of indices passing through that vertex --- the \emph{vertex congestion}.
    Assuming each index is the same size, the tree (c) thus has both a higher maximum contraction width (in bold) and total contraction cost than tree (b).
    }

    \label{fig:ctrees}
\end{figure}

For a given tensor network contraction tree, one can quantify the space and time cost of contracting the network. First, the total space required for the contraction of a network is given, up to an $O(|V|)$ prefactor, by $2^{W}$, for \emph{contraction width}
\begin{equation}
W = \mathrm{ec}_{\mathrm{max}}(B,S) \,,
\end{equation}
where $\mathrm{ec}_{\mathrm{max}}$ is the \emph{maximum edge congestion} for this tree embedding of $G$~\cite{o2019parameterization}. In our notation,
\begin{equation}
\mathrm{ec}_{\mathrm{max}}(B,S) = \max_{v\in V_B} \sum_{e\in s_v} \log_2 w(e) \,.\label{eq:edgecon}
\end{equation}
A \emph{space-optimal contraction tree} for $G$ is then defined by
\begin{equation}
B_{\mathrm{space}}(G) = \underset{B \in \mathcal{B}_{|V|}}{\mathrm{argmin}} \, \mathrm{ec}_{\mathrm{max}}(B,S) \,,\label{eq:opt-space}
\end{equation}
where $\mathcal{B}_{|V|}$ is the set of all rooted binary trees with $|V|$ leaves.
For systems of boolean variables or qubits, $w=2$ and $\mathrm{ec}_{\mathrm{max}}(B,S) = \max_{v\in V_B} |s_v|$. The contraction width is then equal to the maximum vertex degree in the minors of $G$ obtained throughout the contraction path represented by $B$~\cite{Kourtis2018}, as illustrated in the example of Fig.~\ref{fig:ctrees}. The same logic extends to any constant $w$.

Similarly, the time complexity of the contraction is captured by the \emph{contraction cost}
\begin{equation}
C(B,S) = \sum_{v\in V_B} 2^{\mathrm{vc}(B,S,v)} \,,
\end{equation}
where $\mathrm{vc}$ is the \emph{vertex congestion}~\cite{o2019parameterization}
\begin{equation}
\mathrm{vc}(B,S,v) = \sum_{e\in s_{l(v)} \cup s_{r(v)}} \log_2 w(e) \,.\label{eq:vertexcon}
\end{equation}
Again using the case of qubits as an example, the number of operations required to obtain the tensor corresponding to a non-leaf vertex $v$ by contracting its children is proportional to $2^{|s_{l(v)} \cup s_{r(v)}|}$. More precisely, assuming every contraction is an inner product, for real (complex) tensors, the associated FLOP count will be a factor of two (eight) times more than $C$: one (six) FLOP(s) for the multiplication and one (two) FLOP(s) for the addition.
A \emph{time-optimal contraction tree} for $G$ is then
\begin{equation}
B_{\mathrm{time}}(G) = \underset{B \in \mathcal{B}_{|V|}}{\mathrm{argmin}} \, C(B,S) \,.\label{eq:opt-time}
\end{equation}
$B_{\mathrm{time}}(G)$ and $B_{\mathrm{space}}(G)$ are not necessarily the same and hence a strategy that aims to find one is not guaranteed to also find or approximate the other.

\section{Tensor network contraction path optimization}\label{sec:algos}

We have shown that the optimization of the contraction path for a given tensor network corresponds to minimization of a vertex or edge congestion measure over the possible tree embeddings of the network. Instead of performing this minimization, here we will use methods that optimize contraction paths based on quantities that are proxies to these congestion measures, as explained below. Our heuristics are based on established algorithms for a variety of common graph theoretic tasks, such as balanced bipartitioning or community detection, some of which, unlike tree embedding, have seen decades of development and improvement, thus affording great benefits in performance to our methods. We stress, however, that all contraction path optimization tools studied in this work \emph{except} for those introduced in Secs.~\ref{sec:opt} and~\ref{sec:line-graph} are original contributions, and that graph theory algorithms used to perform a particular task (e.g., graph partitioning) are interchangeable with any other algorithm that can perform the same task. Finally, we also note that all the algorithms we test except for the exhaustive search of Sec.~\ref{sec:opt} are not guaranteed to find the global minimum of the congestion measures. Nevertheless, as will be seen below, they can often get arbitrarily close to the optimum. A summary of the methods we introduce below is shown in Tab.~\ref{tab:methods}

\begin{table}[t]
\begin{center}
\resizebox{\columnwidth}{!}{
\begin{tabular}{| c | c | c | c | c |}
 \toprule
Method & Optimal & Edge weights & Hyper edges & Targets \\
\hline\hline
Exhaustive search & yes & yes & yes & total cost\\
\hline
Line graph tree decomposition & depends\footnote{
\texttt{QuickBB} will eventually find the optimal contraction with respect to leading cost but not \texttt{FlowCutter}.} & no & yes & leading cost\\
\hline
Community detection & no & yes  & no & total cost\\
\hline
Boltzmann-greedy & no & yes & yes & total cost\\
\hline
Hyper-graph partitioning & no & yes & yes & total cost\\
\hline
\end{tabular}
}
\end{center}
\caption{\label{tab:methods} Contraction path optimization methods detailed in Secs.~\ref{sec:opt}-\ref{sec:hyper}. For each method, we list its name, whether it is guaranteed to find the optimal contraction path, whether it incorporates edge weights (i.e., bond dimensions), whether it naturally handles hyper-edges, and whether it targets the total contraction cost or just the leading cost (single most expensive contraction).
}
\end{table}

\subsection{Exhaustive search}\label{sec:opt}

One method for finding contraction trees is to exhaustively search through all of them and return whichever minimizes the desired target $W$ or $C$.
Since outer products are rarely ever beneficial, an efficient but virtually optimal way to perform this search is to adopt a dynamic programming approach that builds the tree considering connected subgraphs only~\cite{Pfeifer2014}.
We refer to this optimizer as \texttt{Optimal} and for our results use the version implemented in \texttt{opt\_einsum}~\cite{Smith2018}.

\subsection{Line-Graph Tree Decompositions - \texttt{QuickBB} \& \texttt{FlowCutter}} \label{sec:line-graph}

The most common approach to contracting arbitrary tensor networks in recent years, motivated by the results of Markov and Shi~\cite{Markov2008}, has been to find a tree decomposition of the line graph of $G$.
From this tree decomposition, an edge elimination ordering can be constructed such that the complexity of the corresponding contraction is upper bounded by the tree-width of the line-graph minus one.
Practically speaking, we turn an edge ordering, $(e_1, e_2, e_3, \ldots )$ into a contraction tree as follows.
First, find the subgraph of $G$ induced by the next edge in the ordering, $e_i$.
Update $G$ by contracting all of the tensors in this subgraph to form a single vertex (if there are more than 2 tensors use an exhaustive or greedy approach to find a contraction sequence for this small subgraph only).
Repeat until all edges in the ordering have been processed.

In the tensor network literature the most commonly used tree decomposition finder is \texttt{QuickBB}~\cite{gogate2004complete}, which implements a depth-first `branch and bound' search.
Broadly speaking this approach emphasizes performance for graphs with modest numbers of edges, where indeed \texttt{QuickBB} has been shown to work well~\cite{Dumitrescu2018}.
More recently, the \texttt{FlowCutter} tree decomposition finder~\cite{hamann2018graph,Strasser2017}, has been applied to tensor networks~\cite{Dudek2019}.
\texttt{FlowCutter} takes more of a `top-down' approach which emphasizes performance on graphs with large numbers of edges.
Both function as `any-time' algorithms, able to yield the best found solution after setting an arbitrary time.
On the other hand, neither of these optimizers take edge weights into account, which may be a significant disadvantage in the many-body setting, where, unlike in quantum circuits, bond dimensions can vary significantly.

\subsection{Community detection via edge betweenness - \texttt{Hyper-GN}} \label{sec:community}

One of the methods for the contraction of tensor networks with arbitrary structure introduced in Ref.~\cite{Kourtis2018} is based on detecting communities in the network. Qualitatively, a community is a subset of the vertices in a network that is densely connected internally and sparsely connected with its complement. Detecting communities in networks is a central problem in the study of complex networks~\cite{Porter2009a,Fortunato2010}.

The intuition behind using the community structure to contract an arbitrary tensor network is that it is advantageous to contract all the edges between vertices that belong to a community \emph{first}. That is because the vertex that results from the contraction of all edges within a community, which we call a community vertex, is sparsely connected with the rest of the network. Thus, when a community structure exists and is detected in the network, the adherence of contractions to this community structure is expected to lead to community vertices with a maximum degree that is lower than that of the same number of vertices reached by an arbitrary sequence of contractions of the original network. This approach hence effectively minimizes the contraction cost, i.e., yields a contraction sequence that approximates the one defined by the space-optimal contraction tree.

A popular community structure detection algorithm is the one of Girvan and Newman~\cite{Girvan2002}. It operates by evaluating a quantity called \emph{edge betweenness centrality}, defined as
\begin{equation}\label{eq:edge-centrality}
g(e) = \sum_{s,t\in V} \sigma_{st}(e) / \sigma_{st} \,,
\end{equation}
where $\sigma_{st}$ is the total number of shortest paths between vertices $s$ and $t$, and $\sigma_{st}(e)$ is the number of those paths that pass through edge $e\in E$. The algorithm starts with an empty edge list and repeats two steps:
\begin{enumerate}[leftmargin=2em,itemsep=5pt,parsep=0pt,topsep=1em]
\item remove $e' = \underset{e\in E}{\mathrm{argmax}} \ g(e)$ from $E$ and add it to the list,
\item calculate $g(e) \ \forall \ e \in E$,
\end{enumerate}
until exhausting $E$. Multiple edges can be processed simultaneously, since they have the same $g$. The resulting list of edges, sometimes called a \emph{dendrogram}, defines the detected community structure: if one sequentially removes the list entries from $E$ until $G$ becomes disconnected, then the resulting connected components are the communities of $G$. The algorithm then proceeds by splitting each connected component into smaller communities, and the process repeats all the way down to the individual vertex level.

The output of the Girvan-Newman method is also a contraction path: one simply has to traverse the edge list in reverse, each entry defining a contraction of the endpoints of the corresponding edge.
One can incorporate edge weights (and thus bond dimensions) into Eq.~\eqref{eq:edge-centrality}, possibly randomized with some strength $\tau$, to generate varied paths.
We call the optimizer based on repeated sampling of these paths \texttt{Hyper-GN}.

\subsection{Agglomerative contraction trees - \texttt{Hyper-Greedy}} \label{sec:greedy}

One simple way to construct a contraction tree is greedily from the bottom up.
Here, one ignores any overall structure of the graph $G$ and instead  heuristically scores each possible pairwise contraction.
Based on these scores, a pair of tensors can be chosen and contracted into a new vertex and the list of scores then updated with any new possible contractions.
Whilst we know the exact cost and output size of each pairwise contraction, we do not know the effect it might have on the cost and size of later contractions, meaning we must instead carefully choose the heuristic score function.

Given two tensors $T_i$ and $T_j$ whose contraction yields $T_{k}$, one reasonable choice for the heuristic cost function is
\begin{equation} \label{eq:greedy-cost-function}
    \mathrm{cost}(T_i, T_j) =
    \mathrm{size}(T_{k}) -
    \alpha (\mathrm{size}(T_{i}) + \mathrm{size}(T_{j}))
\end{equation}
with $\alpha$ a tunable constant.
If we take $\alpha=1$ then this cost is directly proportional to the change in memory should we perform the contraction.
Whereas instead taking $\alpha=0$ essentially just prioritizes the rank of the new tensor.
Since we will want to sample many greedy paths we also introduce a `Boltzmann factor' weighting of the costs such that the probability of selecting a pairwise contraction is
\begin{equation} \label{eq:greedy-boltzmann-factor}
    p(T_i, T_j) \propto \exp{(-\mathrm{cost}(T_i, T_j) / \tau)} \,,
\end{equation}
with $\tau$ an effective temperature governing how `adventurous' the path finding should be.
Repeatedly generating contraction trees using this combination of cost and weighting, whilst potentially tuning both $\alpha$ and $\tau$, leads to the \texttt{Hyper-Greedy} optimizer.
\texttt{Hyper-Greedy} generally outperforms other greedy approaches and is quick to run, making it a simple but useful reference algorithm.

\subsection{Divisive contraction trees - \texttt{Hyper-Par}}\label{sec:hyper}

The greedy or agglomerative approach is a natural way to think about building contraction trees from the bottom up.
However, as introduced in \cite{Kourtis2018} we can also try and build contraction trees from the top down in a \emph{divisive} manner.
The key here is that each node in a contraction tree represents not only an effective tensor but a subgraph of the initial graph describing the full tensor network.
As we ascend a contraction tree, merging two nodes corresponds to a pairwise contraction of the two effective tensors.
In reverse, as we descend a contraction tree, splitting a node corresponds to a bipartitioning of subgraph associated with that node.

Practically we start with the list of `childless' vertices - initially just the root of the tree corresponding to the full graph, $\{V_G\}$.
We take the next childless vertex, $V$, and partition it into $V = V_1 \cup V_2$. If $|V_1| > 1$ we append it to the list of childless vertices and similarly if $|V_2| > 1$.
This process can be repeated until the full contraction tree is generated.
Such a divisive approach is very similar to the community detection scheme introduced earlier, however, whilst the Girvan-Newman algorithm naturally yields the entire contraction tree, here we create single contractions one at a time.
This allows one to combine partitioning with other optimizers.
For example, we can instead partition a vertex $V$ into $k$ partitions, $V_1, V_2, \ldots, V_k$ and then use the \texttt{Optimal} or \texttt{Hyper-Greedy} optimizer to `fill in' the contraction tree --- essentially find the contraction path for a tensor network composed just of the tensors corresponding to each of these new subgraphs.
Similarly, if the size of $V$ drops below some threshold, we can again use either \texttt{Optimal} or \texttt{Hyper-Greedy} to find the remaining part of the contraction tree corresponding just to the leaf tensors in $V$.

The cost of an individual contraction - a vertex bi-partitioning - is given by the product of the dimensions of the involved indices.
These include any outer indices of the subgraph, plus any indices that cross the newly created partition.
Since the outer indices are independent of the partition, minimizing the number of indices cut by a partition also minimizes the cost of the corresponding contraction.
This is still essentially a greedy approach - it only considers the cost of a single contraction and strictly minimizing this cost (corresponding to choosing a min-cut) could likely create more expensive contractions down the line.
However, one way to heuristically adjust this is to control how \emph{balanced} to make the partitions, in other words, how much to match the size of each partition.
Specifically, we can define the \emph{imbalance} parameter, $\epsilon$, such that $|V_i| \leq (1 + \epsilon) |V| / k$ for $i=1 \ldots k$, where $k$ is the number of partitions.
If $\epsilon$ is close to zero, then the partitions are forced to be very similar in size, whilst if $\epsilon$ is close to $k$ the partitions are allowed to be of any size.

 \begin{figure}[t]
     \centering
     \includegraphics[width=0.99\columnwidth]{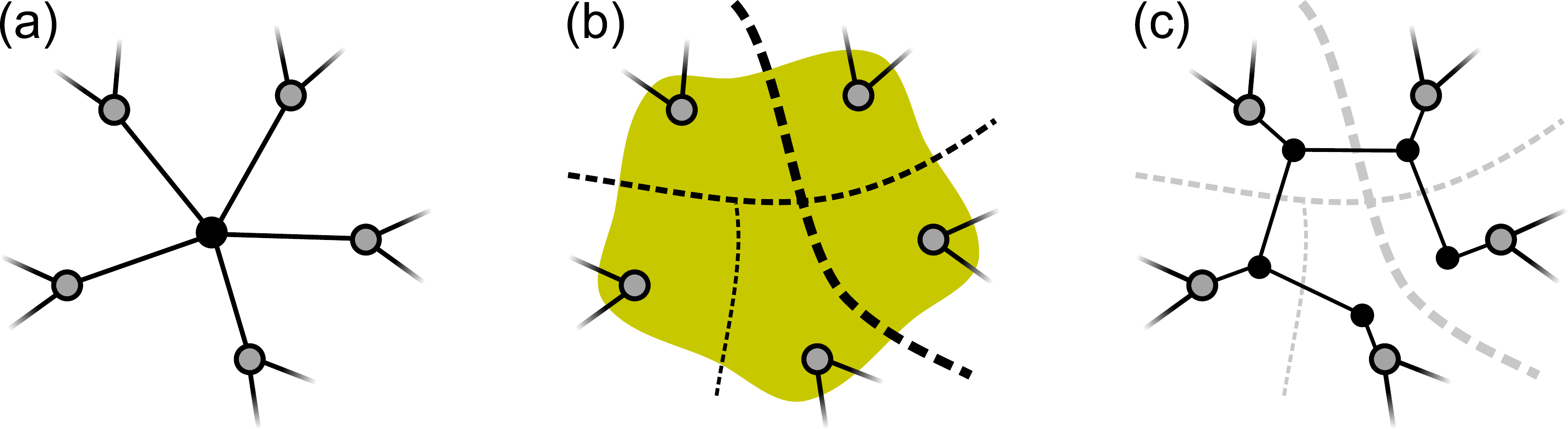}
     \caption{(a) Segment of tensor network with six tensors, one of which (black filled circle) is a COPY tensor. (b) COPY tensor replaced by a hyperedge. Recursive hypergraph bipartitioning yields the separator hierarchy drawn as dashed lines, with thicker lines for higher level in the hierarchy. (c) After a separator hierarchy is found, the hyperedge is replaced by a connected subgraph of COPY tensors whose edges intersect each separator at most once. The results of the contraction of networks (a) and (c) are identical.}
     \label{fig:hyper}
 \end{figure}

Taking into account the internal structure of the tensors in a problem allows for further flexibility in the recursive bipartition process, which in turn can lead to significant performance gains. As an example, consider the case of a COPY tensor, whose entries are 1 only when all indices are equal and 0 otherwise. These tensors appear, for example, when modeling circuits of controlled gates (see, e.g., Sec.~\ref{sec:gate-decomp}) or satisfiability formulas~\cite{Denny2012, Kourtis2018}. Each COPY tensor in a network can be replaced by any connected graph of COPY tensors without changing the result of the contraction~\cite{Biamonte2017}. By replacing all COPY tensors in the network with \emph{hyperedges}, one can perform recursive \emph{hypergraph} bipartitioning with more freedom in the search for short cuts compared to the original graph. To revert back to a `traditional' tensor network after partitioning, each hyperedge can be replaced by a low-rank COPY tensor subgraph that cuts each separator at most once, as illustrated in Fig.~\ref{fig:hyper}.
Another important use-case for hyperedges is to efficiently treat batch and output indices, though these are not benchmarked in this work.

We employ the partitioner KaHyPar~\cite{Schlag2016,Akhremtsev2017} to generate our contraction trees for a number of reasons.
Aside from offering state-of-the-art performance, it also can handle hypergraphs (and thus arbitrary tensor expressions), allows key parameters such as the imbalance to be specified, and takes into account edge weights (and thus arbitrary bond dimensions).
Repeatedly sampling contraction trees whilst tuning the parameters $k$, $\epsilon$ and the cut-off to stop partitioning leads us to the optimizer we call \texttt{Hyper-Par}. Note that the line graph and greedy methods of Secs.~\ref{sec:line-graph} and~\ref{sec:greedy}, respectively, also support hypergraphs natively.

In passing, we note that (hyper)graph partitioning has been used as a simplification tool for computational tasks in other research fields, see, e.g.,~\cite{Papa2007hypergraphpartitioning}.

\subsection{Stochastic Bayesian Optimization}

The \texttt{Optimal} contraction tree optimizer runs until completion whilst \texttt{QuickBB} and \texttt{FlowCutter} are natively any-time algorithms. For the remaining three optimizers -- \texttt{Hyper-GN}, \texttt{Hyper-Greedy} and \texttt{Hyper-Par} -- we use a combination of randomization and Bayesian optimization~\cite{shahriari2015taking} to intelligently sample ever better contraction paths.
This allows all three of them to run as parallel any-time algorithms.

For the \texttt{Hyper-GN} and \texttt{Hyper-Par} optimizers, randomization can be introduced as a noise of the edge weights of the initial graph $G$. For the \texttt{Hyper-Greedy} optimizer the Boltzmann sampling of greedy contractions yields another source of randomization.
Due to the high sensitivity of the contraction width $W$ and cost $C$ to the contraction path, simply sampling many paths and keeping the best already offers significant improvements over single shot versions of these same algorithms.
However we can further improve the performance if we allow the heuristic parameters of each optimizer to be tuned as the sampling continues.
We use the \texttt{baytune}~\cite{Laura:2018} library to perform this optimization, which uses \emph{Gaussian processes}~\cite{williams2006gaussian} to model the effect of the parameters on the target score -- either $W$ or $C$ -- and suggest new combinations which are likely to perform well.

\subsection{Tensor Network Simplifications}
\label{sec:simplification}

Next we describe a series of simplifications based simply on tensor network structure and sparsity of the tensors that we perform iteratively until no more operations are possible.
These are all designed to decrease the complexity of the tensor network prior to invoking the full contraction path finders, and are performed as efficient local searches.

The first of these is \emph{diagonal-reduction} of tensor axes, as introduced for quantum circuits in~\cite{Boixo2017}.
For a $k$-dimensional tensor, $t_{i_1 i_2 \ldots i_k}$, with indices $i_1 i_2 \ldots i_k$, if for any pair $\{i_x, i_y\}$
\begin{equation}
    t_{i_1 i_2 \ldots i_k} = 0 ~~\forall~~ i_x \neq i_y
\end{equation}
then we can replace $t$ with a $(k-1)$-dimensional tensor, $\tilde{t}$ with elements $\tilde{t}_{\ldots i_x} = t_{\ldots i_x i_y} \delta^{i_x}_{i_y}$, where the $\delta$ copy can be implemented by re-indexing $i_y \rightarrow i_x$ everywhere else in the tensor network, thus resulting in $i_x$ becoming a \emph{hyperedge}. This enables the use of the hypergraph machinery detailed in Sec.~\ref{sec:hyper}.

\begin{center}
\includegraphics[width=0.9\linewidth]{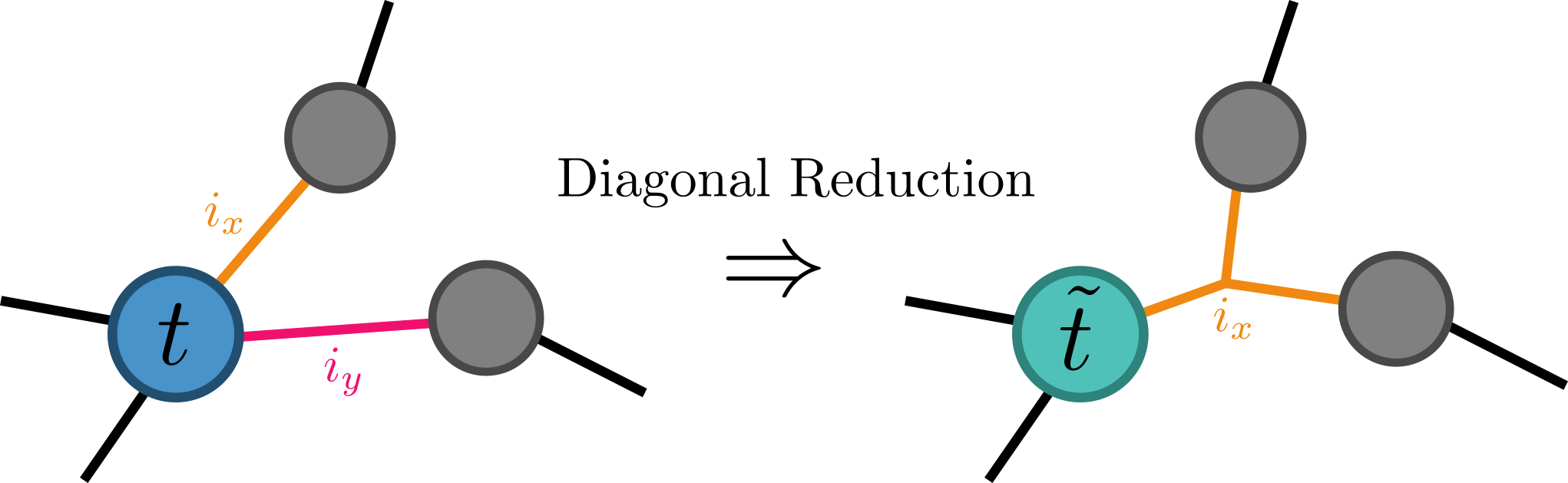}
\end{center}

The second pre-processing step we perform is \emph{rank-simplification}.
Here we generate a greedy contraction path that targets rank reduction only (i.e. with respect to Eq.~\eqref{eq:greedy-cost-function} and \eqref{eq:greedy-boltzmann-factor} sets $\alpha=\tau=0$).
We then perform any of the pairwise contractions such that the rank of the output tensor is not larger than the rank of either input tensor.
If the tensor network has no hyperedges, this corresponds to absorbing all rank-1 and rank-2 tensors into neighbouring tensors, a process which cannot increase the cut-weight across any partition for example.

\begin{center}
\includegraphics[width=0.9\linewidth]{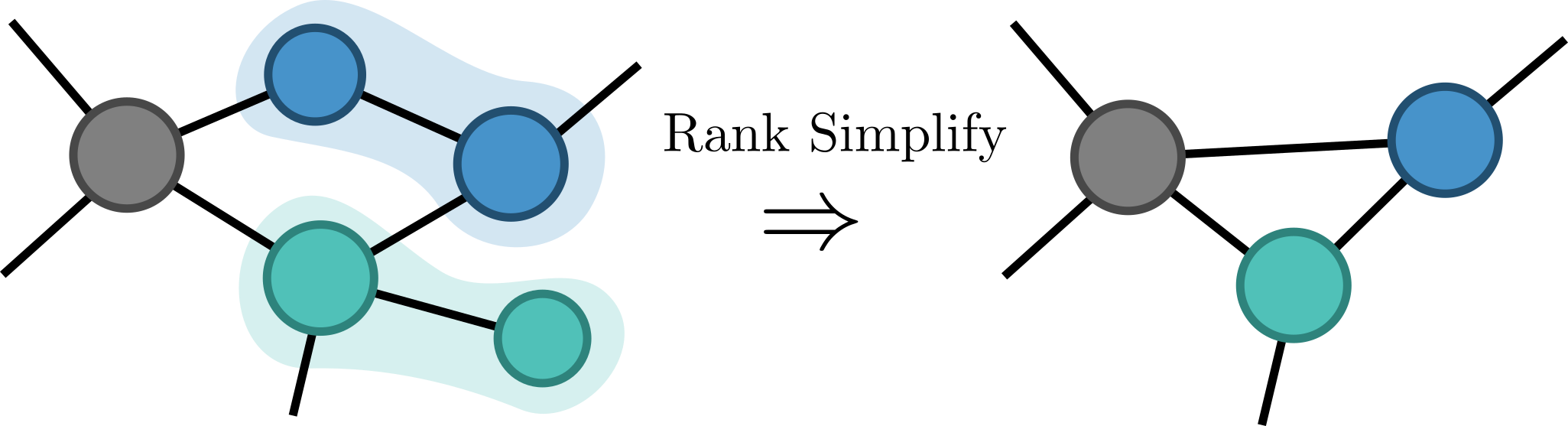}
\end{center}

The third pre-processing step we perform is \emph{antidiagonal-gauging}.
Here, again assuming we have a $k$-dimensional tensor $t_{i_1 i_2 \ldots i_k}$, if for any pair of indices $\{i_x, i_y\}$ of matching size $d$ we find
\begin{equation}
    t_{i_1 i_2 \ldots i_k} = 0 ~~\forall~~ i_x \neq d - i_y
\end{equation}
then we can flip the order of either index $i_x$ or $i_y$ throughout the tensor network.
This corresponds to gauging that index with a `reflected' identity, for example if $d=2$ the Pauli matrix $X$.
This simplification does not help on its own but merely produces tensors which can then be diagonally reduced using the prior scheme.

\begin{center}
\includegraphics[width=0.9\linewidth]{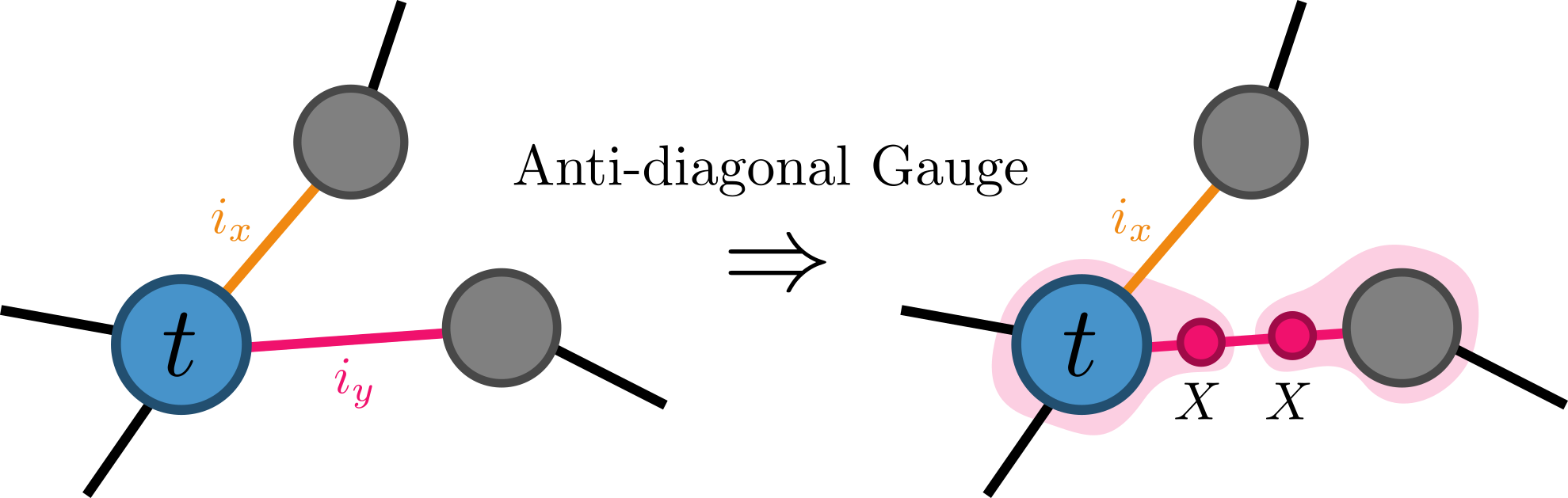}
\end{center}

The fourth simplification we perform is \emph{column-reduction}.
Here, if for any $k$-dimensional tensor $t_{i_1 i_2 \ldots i_k}$ we find an index $i_x$ and `column' $c$ such that
\begin{equation}
    t_{i_1 i_2 \ldots i_k} = 0 ~~\forall~~ i_x \neq c
\end{equation}
then we can replace every tensor, $t_{\ldots i_x}$, featuring that index with the $(k-1)$-dimensional tensor $\tilde{t}$ corresponding to the slice $t_{\ldots[i_x=c]}$, removing that index from the network entirely.
This can be pictured as projecting the index into the basis state $\ket{c}$.

\begin{center}
\includegraphics[width=0.9\linewidth]{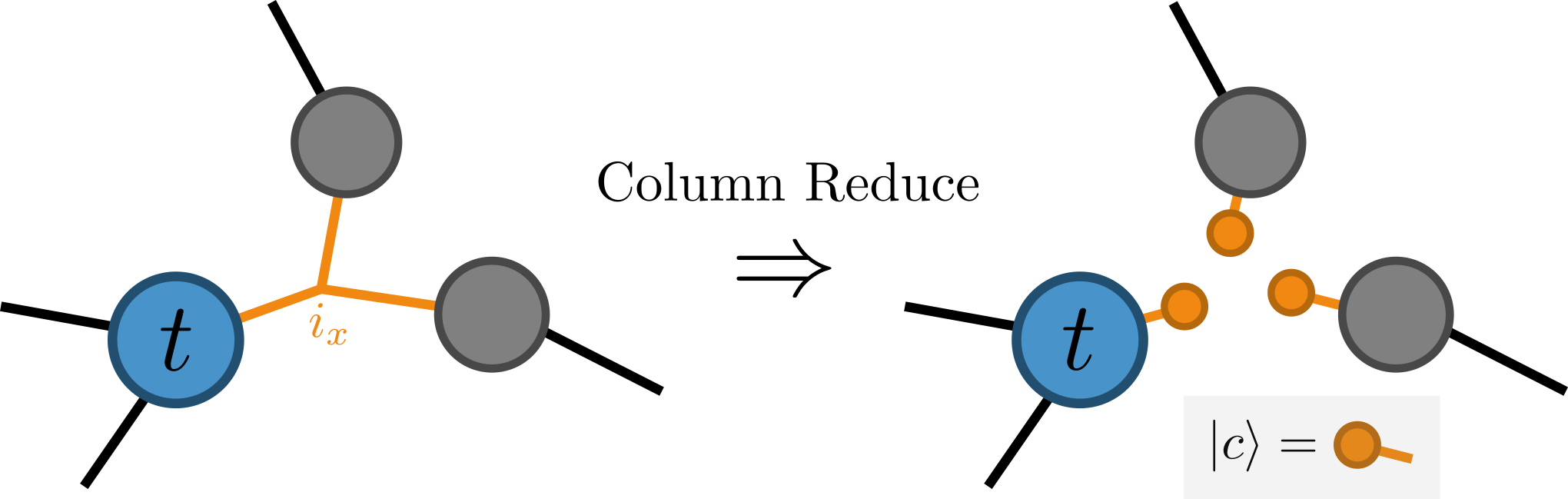}
\end{center}

The final possible processing step is \emph{split-simplification}.
Here if any tensor, $t$, has an exact low-rank decomposition across any bipartition of its indices -- i.e.
$
t_{i_1 \ldots j_1 \ldots} = \sum_k l_{i_1 \ldots, k} r_{j_1, \ldots, k}
$ with
$
\max(\mathrm{size}(l), \mathrm{size}(r)) < \mathrm{size}(t)
$ -- we perform it.
This is done using the SVD, and is the one simplification that increases the number of tensors in order to decrease the cut-weight across partitions.

\begin{center}
\includegraphics[width=0.9\linewidth]{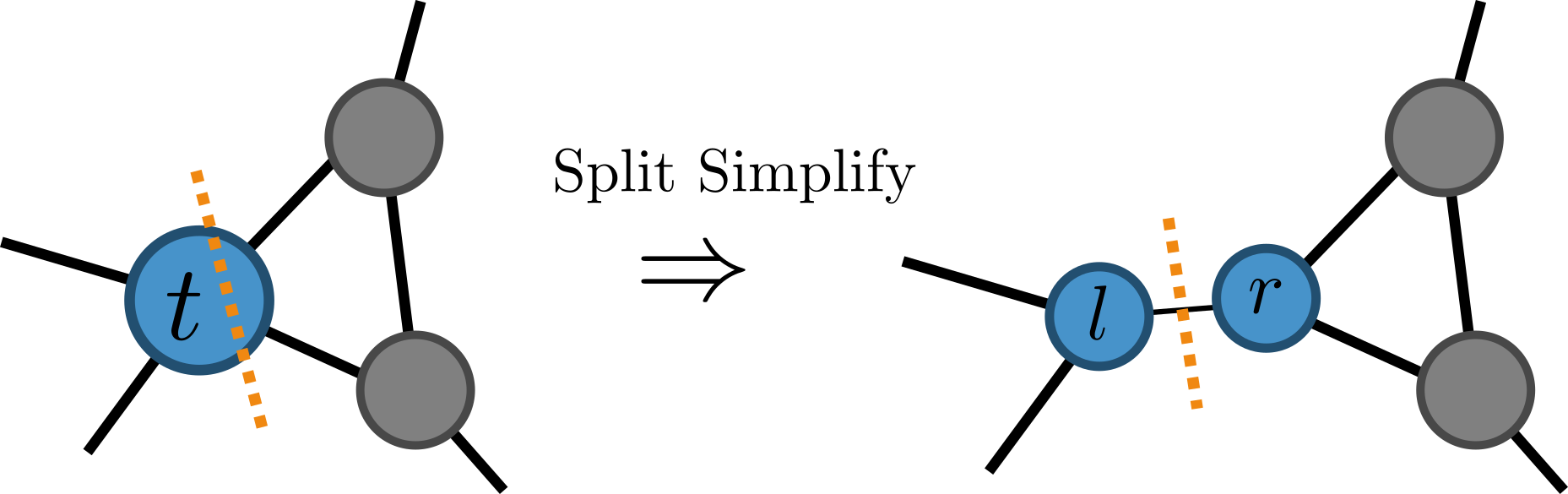}
\end{center}

We apply the above set of simplifications iteratively but deterministically until no method can find any operation to perform.
For all methods that compare to zero we use a relative precision of $10^{-12}$ unless otherwise stated.
The order they are applied in can produce very different networks -- we find cycling through the order \{\emph{antidiagonal-gauging, diagonal-reduction, column-reduction, rank-simplification, split-simplification}\} produces good results.
Indeed for quantum circuits generally the resulting tensor networks often have almost no sparsity among tensor entries.
Note for methods such as \texttt{Hyper-GN} which cannot handle hyperedges we skip the diagonal-reduction.
Finally, if aiming to reuse a contraction path, one needs to maintain the sparsity structure from network to network, possibly excluding any variable tensors from the simplification steps that detect sparsity.
For most circuits terminated with a layer of Hadamard gates, if one only changes the sampled bit-string $x$ then even this is not usually necessary.

\begin{figure*}[tb]
    \includegraphics[width=\textwidth]{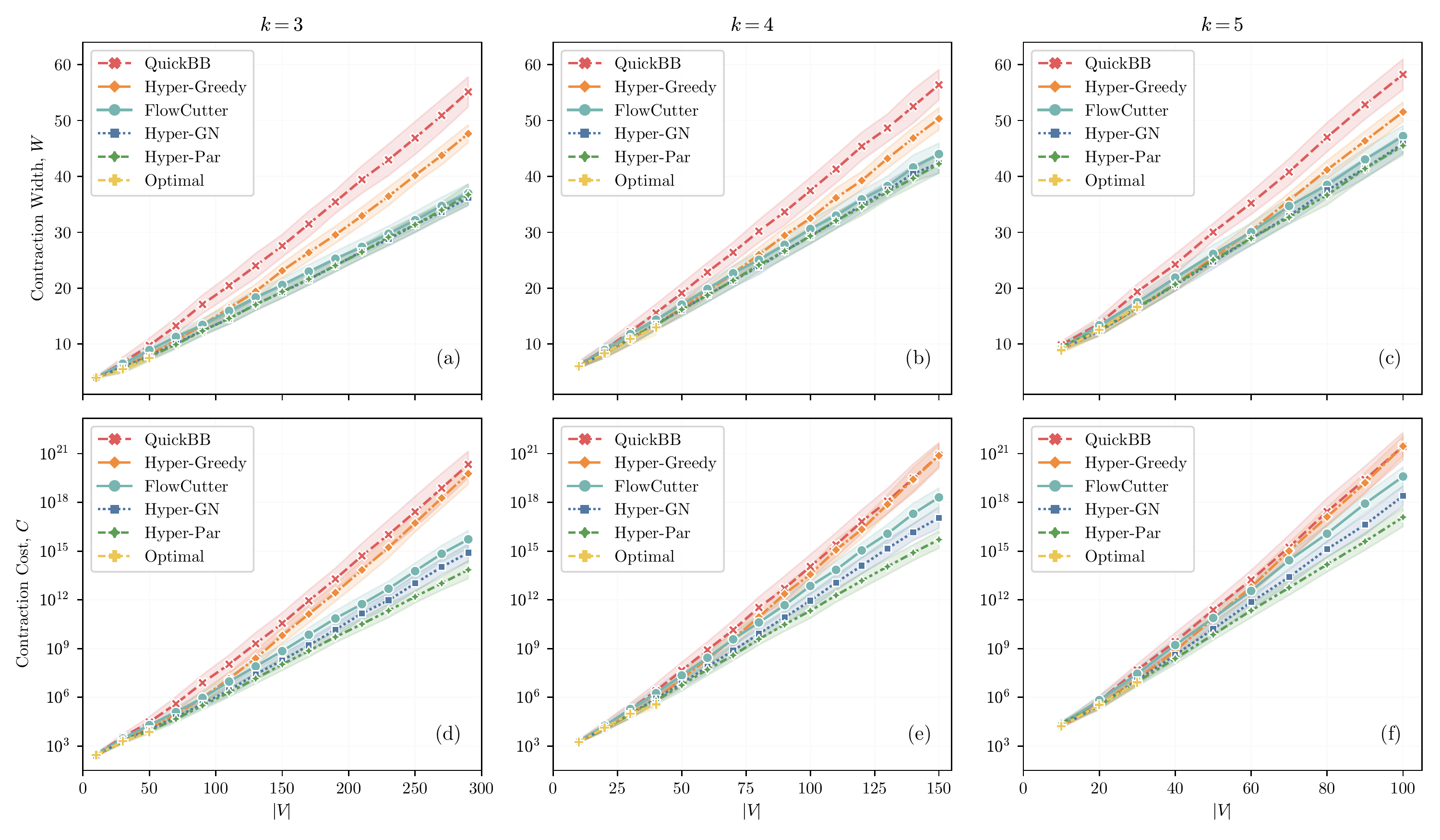}
    \caption{
        Mean contraction width (top row) and cost (bottom row) of random regular graphs of degree $k=3, 4, 5$ (left, centre and right columns respectively) as a function of the number of vertices (tensors) in the network, $|V|$, for various contraction path optimizers each allowed 5 minutes to search.
        The shaded regions show standard deviations across 100 random graph instances.
        An example graph with $k=5$ is shown in Fig.~\ref{fig:pic-graphs}(b).
    }
    \label{fig:rand-reg}
\end{figure*}

\section{Results}\label{sec:results}

We benchmark our contractors on six classes of tensor networks with complex geometry -- random regular graphs, random planar graphs, square lattices, weighted model counting formulae, QAOA energy computation, and random quantum circuits.
In each set of results we set a time limit or maximum number of shots for each of the optimizers to run for, and then target either the contraction width, $W$, or contraction cost $C$.
As a reminder, $W$ is essentially the space requirement of the contraction ($\log_2$ of the size of the largest intermediate tensor) whilst $C$ is the time requirement (the total number of scalar operations).
The \texttt{Optimal} algorithm is able to search for either the minimum $W$ or $C$, whilst \texttt{Hyper-GN}, \texttt{Hyper-Greedy} and \texttt{Hyper-Par} can target either through the guided Bayesian optimization.
Finally, there is no way to specifically bias \texttt{QuickBB} and \texttt{FlowCutter} towards either $W$ or $C$ so in each case the optimizer runs identically.
If an optimizer can run in parallel, we allow it 4 cores to do so.
An open source implementation of the optimizers, compatible with \texttt{opt\_einsum}~\cite{Smith2018} and \texttt{quimb}~\cite{gray2018quimb}, is available at~\cite{cotengra}.

To give some context to the relative scale of $W$ and $C$, a complex, single precision tensor of size $2^{27}$ requires 1GB of memory, and a consumer grade GPU can usually achieve a few teraFLOPs in terms of performance, corresponding to $C \sim 10^{15}$ over an hour.
In the final results section we benchmark various contractions and indeed find this real-world performance.
At the extreme end of the scale, the most powerful supercomputer in the world currently, Summit, has a few petabytes of memory, corresponding \emph{very roughly} to $W \sim 47$, though this is obviously distributed among nodes and utilizing it for a single contraction would need, among many other technical considerations, significant inter-node communication.
Summit has also achieved sustained performance of a few hundred petaFLOPs~\cite{villalonga2019establishing}, which over an hour might correspond to $C\sim 10^{20}$, but is unlikely to do so if distributed contraction is required (i.e. for high $W$).

\subsection{Random Regular Graphs}
\label{sec:rand-reg}

We start by benchmarking tensor networks with geometries defined by random regular graphs, as studied in~\cite{Kourtis2018,Dudek2019}. These graphs arise in the study of many computational problems, such as satisfiability, but also problems defined on graphs with nonuniform degree distribution can often be reduced to equivalent problems on low-degree regular graphs~\cite{Markov_2009}.
For such a $k$-regular graph, every vertex is connected randomly to $k$ others, with total number of vertices $|V|$.
We treat each of the edges as tensor indices of size 2 and associate a rank-$k$ tensor with each vertex.
None of the simplifications of Sec.~\ref{sec:simplification} are applicable.
An example of such a network is shown in Fig.~\ref{fig:pic-graphs}(b).
For each size $|V|$, degree $k$ and target $\in \{W, C\}$, we generate 100 sample regular graphs uniformly~\cite{Viger2005}, and allow 5 minutes of search time per instance for each optimizer.
The reference \texttt{Optimal} path finder we instead run for 24 hours and only show data points where all but one or two of the instances successfully terminated so as not to bias those points towards easy instances.

The results are shown in Figs.~\ref{fig:rand-reg}(a)-(f).
First of all we note that for small sizes all optimizers return similar performance, indeed, close to \texttt{Optimal}.
As $|V|$ increases however the same ranking emerges in each combination of $k$ and $\{W, C\}$: (from worst to best) \texttt{QuickBB}, \texttt{Hyper-Greedy}, \texttt{FlowCutter}, \texttt{Hyper-GN}, then finally \texttt{Hyper-Par}.
We attribute the improvement of \texttt{Hyper-GN} over previous studies~\cite{Dudek2019} to the use of guided stochastic sampling.
There are some interesting performance comparisons when it comes to targeting contraction width $W$ or cost $C$.
For example, while \texttt{Hyper-Greedy} beats \texttt{QuickBB} for width across the board, the results are much closer for contraction cost.
On the other hand, the advantage of \texttt{Hyper-Par} over \texttt{Hyper-GN} and \texttt{FlowCutter} is much more pronounced when considering cost rather than width.

\begin{figure}[tb]
    \centering
    \includegraphics[width=\linewidth]{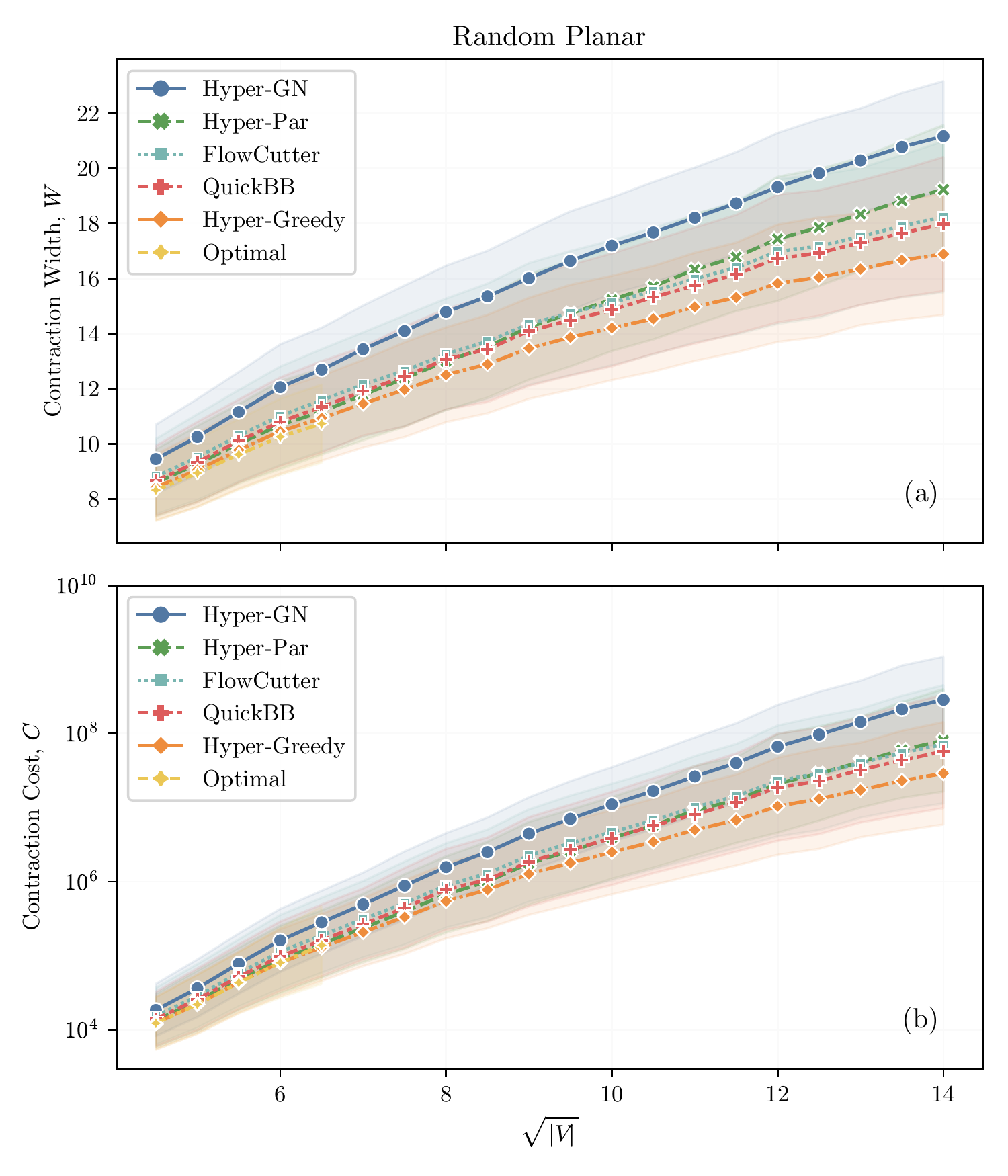}
    \caption{
        Mean contraction width $W$ (top) and cost $C$ (bottom) for randomly generated planar graphs as a function of number of vertices $|V|$, for various path optimizers each allowed 5 minute to search.
        The shaded regions show standard deviations across random graph instances.
        The 35,162 graph instances studied are approximately uniformly distributed over the $\sqrt{|V|}$ bins shown, and an example instance is shown in Fig.~\ref{fig:pic-graphs}(c).
    }
    \label{fig:rand-planar}
\end{figure}

\subsection{Random Planar Graphs}
\label{sec:rand-planar}

A contrasting class of geometries to consider is that of planar graphs, encountered for example in the study of physical systems defined on a 2D lattice or in evaluating knot invariants~\cite{Meichanetzidis2019}.
To investigate these in a generic fashion, we generate random planar graphs with $|V| \in [20, 200]$ using the Boltzmann sampler described in~\cite{Fusy2009}.
An instance of the generated graphs is shown in Fig.~\ref{fig:pic-graphs}(c).
Whilst these are much more random than square lattices for example, we find nonetheless that the results are broadly representative.
Similarly to the random regular graphs, for each vertex with $k$ edges we associate a rank-$k$ tensor with bond dimensions of size 2 and allow each optimizer 5 minutes per instance to explore contraction paths.
In~\cite{Dudek2019} it was shown that the optimal contraction path with respect to $W$ for planar graphs can be found in polynomial time.
Also, planar tensor networks can be contracted in subexponential time $O(2^{\sqrt{|V|}})$ as a consequence of the planar separator theorem~\cite{Cai2007,Valiant2008,Kourtis2018}.
In Fig.~\ref{fig:rand-planar}(a) and (b) we plot the mean contraction width, $W$, and cost, $C$, as a function of the `side length' of the graph, $\sqrt{|V|}$.
Alongside a sub-exponential scaling for all the optimizers we see a very different ranking of optimizer performance as compared to random regular graphs, with \texttt{Hyper-Greedy} performing best.
For small sizes, again the performance of all optimizers is close to \texttt{Optimal}, and in fact the difference between methods remains relatively small throughout the size range.

\subsection{Regular Square Lattice}

\begin{figure}[tb]
    \centering
    \includegraphics[width=\linewidth]{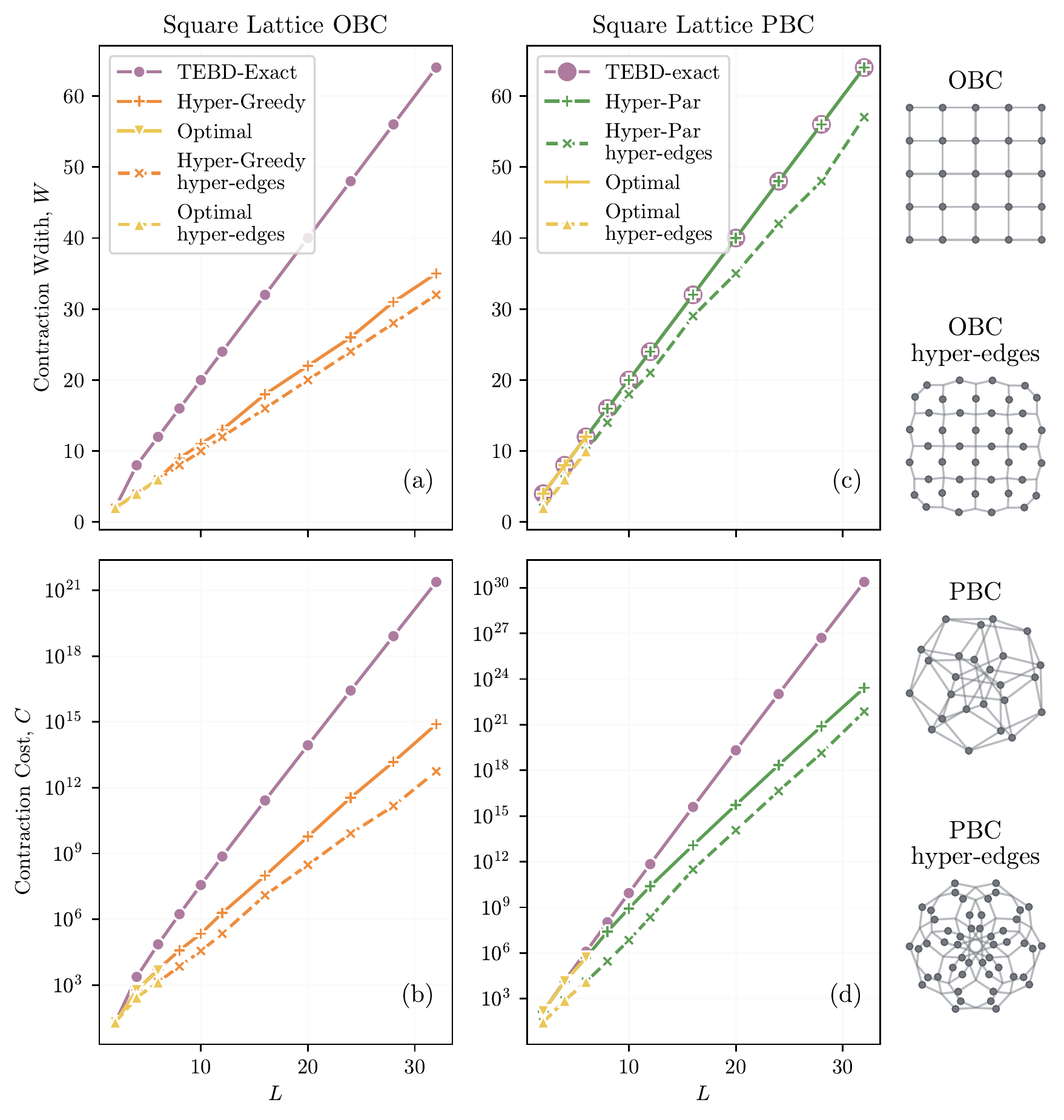}
    \caption{Contraction width $W$ (top row) and contraction cost $C$, for square lattice geometry - either with vertices representing the underlying lattice (left column) or hyper-edges (right column).
    Insets to right illustrate the four possible TNs with $L=5$. Note that the hyper-edge case can be exactly transformed into the normal case but the reverse is not generally true.
    }
    \label{fig:tebd-compare-2d}
\end{figure}

To emphasize that the utility of these optimizers is not restricted to randomly structured graphs, we now compare the best of them with a naive Time Evolving Block Decimation (TEBD) style approach on a square 2D lattice.
While such an approach -- contracting a Matrix Product State boundary from one side to the other -- usually would be combined with canonicalization and compression, doing it exactly yields a natural comparison point for a simple, manually chosen contraction path.
In Fig.~\ref{fig:tebd-compare-2d} we show $W$ and $C$ for such an approach (labelled \texttt{TEBD-Exact}), the best of \texttt{Hyper-Greedy} or \texttt{Hyper-Par}, as well as \texttt{Optimal}, for 2D square lattice TNs with bond dimension 2.
As well as showing open and periodic boundary conditions (OBC and PBC), we show the case for when the lattice geometry is defined on hyper-edges rather than the vertices.
This is a common scenario when evaluating partition functions of classical spin models.
While the hyper-edges can be converted to COPY tensors to yield the standard TN geometry, this makes the TN harder to contract.

For OBC, we find $W$ is significantly reduced from the \texttt{TEBD-Exact} scaling
\footnote{With canonicalization but no compression the scaling would be
$ W \sim L$.}
of $2L$ (Fig.~\ref{fig:tebd-compare-2d}(a)) as well as $C$ (Fig.~\ref{fig:tebd-compare-2d}(b)).
Contracting the hyper-edge form of the TN also yields an advantage for both.
For PBC the \texttt{TEBD-Exact} path yields the same, optimal contraction width (Fig.~\ref{fig:tebd-compare-2d}(c)) but carries a significantly worse scaling contraction cost (Fig.~\ref{fig:tebd-compare-2d}(d)).
Contracting the hyper-edge form of the TN again yields an advantage for both.
In all cases we see either \texttt{Hyper-Greedy} or \texttt{Hyper-Par} very closely tracks the \texttt{Optimal} width and cost at accessible sizes.

\subsection{Exact Weighted Model Counting}

We now move on to exact weighted model counting, an important \#P-complete task, central to problems of inference in graphical models, evaluating partition functions in statistical physics, calculating network reliabilities, and many others~\cite{Bacchus2003,Domshlak2007,Gomes2009ModelC}.
The problem can be cast as computing the following sum:
\begin{equation}\label{eq:WMC}
    x = \sum_{\{v\}}
    \prod_v^{\mathrm{\#vars}} w_{v}
    \prod_{i}^{\mathrm{\#clauses}} C_{\bar{v}_i}
    ~,
\end{equation}
where $\{v\}$ is all combinatorial assignments of every binary variable, $w_v$ is a vector with the `positive' and `negative' weight of variable $v$, and $C_{\bar{v}_i}$ the $i^\mathrm{th}$ clause containing variables $\bar{v}_i$, given by the tensorization of the OR function.
Such an expression can directly be thought of as an hyper tensor network, with tensors (nodes) $w_v$, $C_{\bar{v}_i}$ and tensor indices (hyper-edges) $v$.
Key here is that we directly handle constructing contraction trees for such hyper-graphs, and thus do not need to map Eq.~\eqref{eq:WMC} into a `normal' tensor network form.

\begin{figure}[tb]
    \centering
    \includegraphics[width=\linewidth]{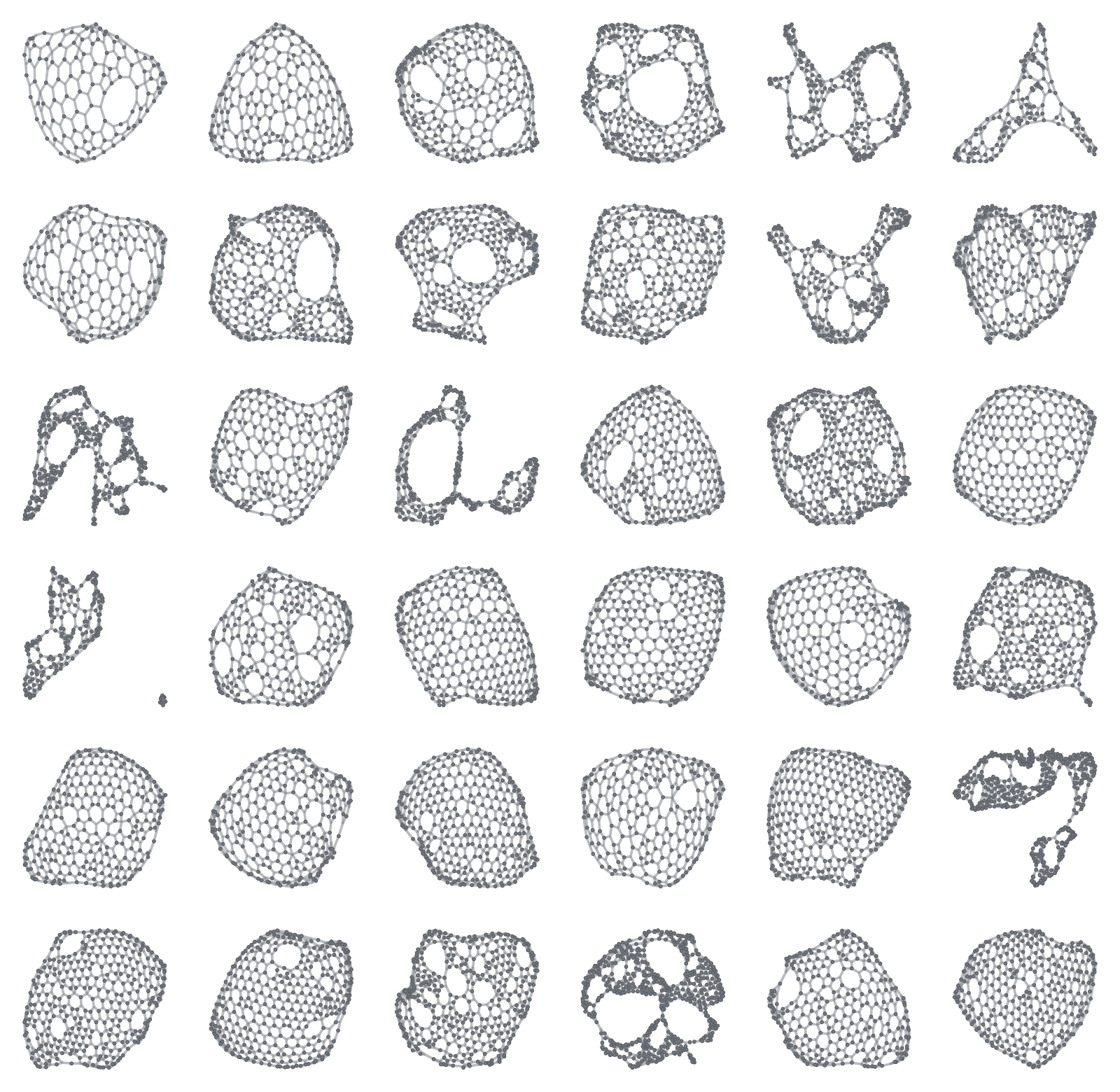}
    \caption{Example hyper tensor networks, post-simplification, representing weighed model counting formulae from the MCC2020 model counting competition.}
    \label{fig:wcnf_plots}
\end{figure}

To test our contraction optimizers we assess all 100 private weighted model counting (track-2) instances from the Model Counting 2020 competition~\cite{mcc2020}.
After constructing the tensor network representation of $x$ we run the simplification procedure, actively renormalizing the tensors since for some instances $x > 10^{2000}$.
We find the simplifications to be very powerful here -- of the 100 instances, 63 simplify all the way to a single scalar, whilst the remaining 37 instances require actual contraction of a much reduced tensor network.
We invoke our hyper-optimizer on these, allowing 64 repeats and access to both the \texttt{greedy} and \texttt{KaHyPar} drivers.
Of these, 1 instance was exceptionally difficult ($W \gtrsim 100$), whilst the remaining (shown in Fig.~\ref{fig:wcnf_plots}) all had contraction paths with $W<20$ and $C<10^{8}$ making them easily contractable.
Overall the 99 solved instances compares favourably with the best score of 69 achieved in the competition~\cite{mcc2020}.
For those 69 instances we confirmed all results against the \texttt{ADDMC} solver~\cite{dudek2020addmc}.

\subsection{QAOA Energy Evaluation}
\label{sec:qaoa}

The Quantum Approximate Optimization Algorithm (QAOA)~\cite{farhi2014quantum} is a promising approach for optimization on near-term quantum devices.
It involves optimizing the energy of an ansatz circuit, followed by the sampling of potential solution bitstrings.
Here we explore the first part, a task that has been studied before~\cite{huang2019alibaba} and is identical to computing the energy of a unitary ansatz for a many-body model.
The $p$-layer ansatz circuit for target graph $G$ with constraint weights $w_{j,k}$ for $j, k\in E(G)$ is given by:
\begin{equation}
    |{\bar{\gamma}, \bar{\beta}}\rangle = U_B (\beta _p)
        U_C (\gamma _p) \cdots U_B (\beta _1) U_C (\gamma _1) |{+}\rangle
\end{equation}
where
\begin{align}
    U_C (\gamma) &= \prod \limits_{j, k\in E(G)} e^{-i \gamma w_{j k} Z_j Z_k}
        \\
    U_B (\beta) &= \prod \limits_{j \in G} e^{-i \beta X_j}
\end{align}
for the two length-$p$ vectors of parameters $\bar{\alpha}$ and $\bar{\beta}$.
The energy of this is given by a sum of local terms:
\begin{equation}
    E = \sum_{j, k\in E(G)}
    w_{j,k}
    \bra{\bar{\gamma}, \bar{\beta}}
    Z_{j} Z_{k}
    \ket{\bar{\gamma}, \bar{\beta}}
\end{equation}
where for each term any unitaries outside the `reverse lightcone' of $j,k$ can be cancelled.

\begin{figure}[tb]
    \centering
    \includegraphics[width=\linewidth]{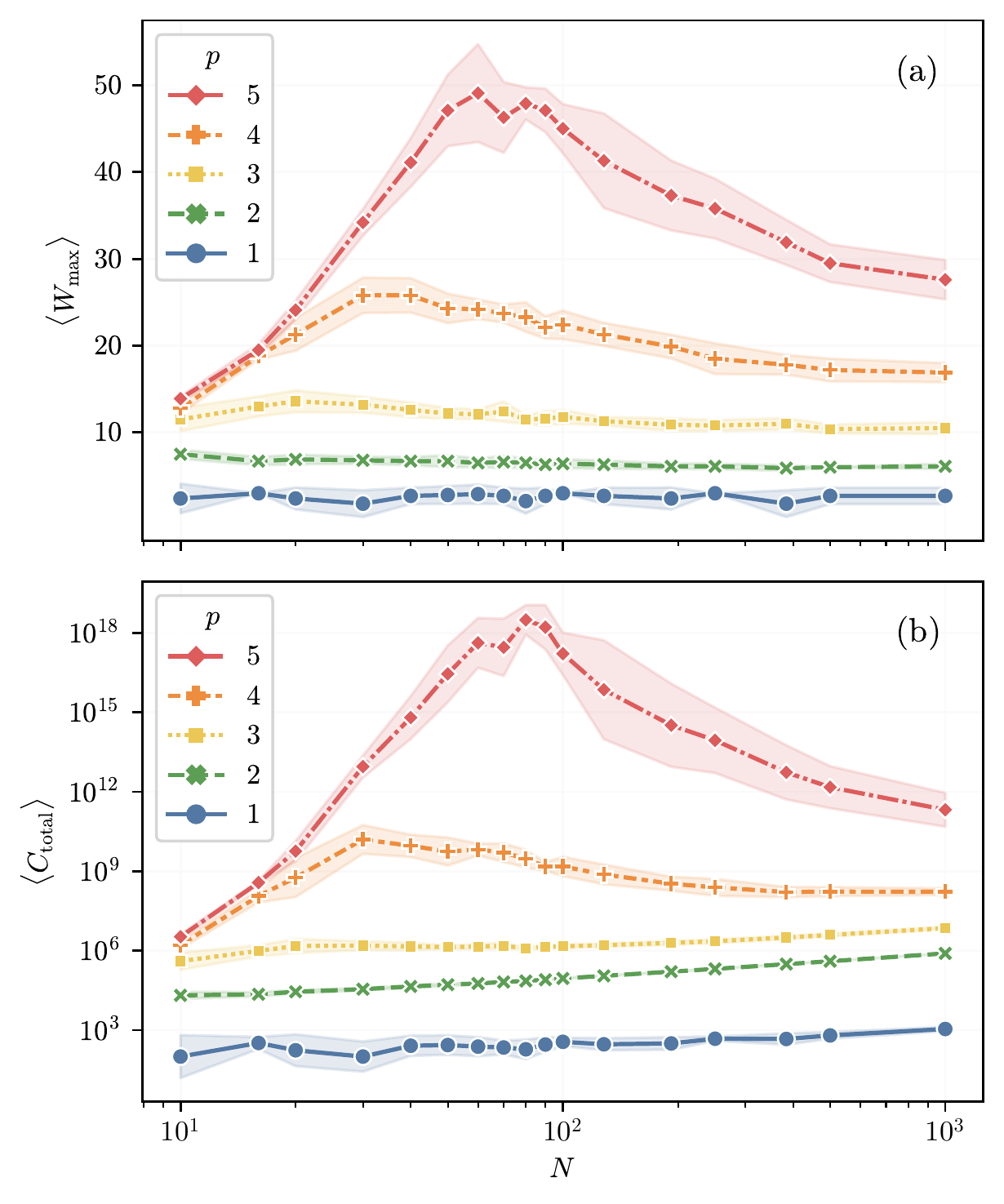}
    \caption{
    Maximum contraction width (a) and total contraction cost (b) for computing the energy of a $p$-layer QAOA circuit, averaged across 10 instances of random 3-regular graphs of size $N$.
    The shaded region shows the standard deviation across these instances.
    }
    \label{fig:qaoa-results}
\end{figure}

We study MAX-CUT problems on random 3-regular graphs of size $N$, for which $w_{j,k}=1$, equivalent to an antiferromagnetic Ising model.
Note that whilst the problem is defined on such a graph, $G$, the actual tensor networks for each energy term have very different geometries compared to Sec.~\ref{sec:rand-reg}, since they arise from the repeated application of $3p$ layers of gates followed by unitary cancellation.
Indeed, in the limit of large $N$, they are not random at all~\cite{huang2019alibaba}.
First we form the $\frac{3N}{2}$ energy term tensor networks, and simplify each using all five methods from Sec.~\ref{sec:simplification}.
We invoke our hyper-optimizer on these, allowing 64 repeats and access to both the \texttt{greedy} and \texttt{KaHyPar} drivers.
In Fig.~\ref{fig:qaoa-results} we report the maximum contraction width, $W_\mathrm{max}$ and total contraction cost, $C_\mathrm{total}$, across terms, averaged over 10 instances of the random regular graphs, as a function of $N$ and $p$.

We note that up to and including $p{=}4$, throughout the range of $N$, $W_\mathrm{max}$ remains less than $\sim 28$ and $C_\mathrm{total}$ less than $\sim 10^{10}$, putting such simulations easily within the range of single workstations.
As an example, on a CPU with 4 cores, performing all of the contractions for $N=54$ and $p=4$ takes on the order of seconds.
Stepping up to $p=5$ increases the difficulty significantly, especially in the $N=40-120$ range.
The peak here is due to cycles of length $\leq p$ appearing in $G$ for small enough $N$, which dramatically increase the complexity of each tensor network.

\subsection{Random Quantum Circuits}

The final class of tensor networks we study is those corresponding to random quantum circuits executed on a range of quantum chip geometries.
In particular, we look at sizes and depths previously explored in the context of so-called `quantum supremacy'~\cite{Preskill2012,Aaronson2016,Boixo2018,Arute2019}.
Quantum circuits can be naturally cast as tensor networks and then simulated via contraction, as shown in~\cite{Markov2008}.
In recent years, random quantum circuits have been used both as a test-bed for tensor network contraction schemes as well as setting the benchmark for demonstrating `quantum supremacy'~\cite{Pednault2017,Boixo2017,Chen2018,Zhang2019,Villalonga2019,guo2019general}.
Practically speaking, such simulations can also allow the fidelity of real quantum chips to be benchmarked and calibrated~\cite{Boixo2018,Villalonga2019,Arute2019}.

The simplest quantity to compute here is the `transition amplitude' of one computational basis state to another through a unitary describing the quantum circuit.
Assuming we start with the $N$ qubit all-zero bit-string $\ket{0^{\otimes N}}$, the transition amplitude for output bit-string $x$ can be written:
\begin{equation} \label{eq:trans-amp}
    c_x = \bra{x} U_d U_{d - 1} \ldots U_2 U_1 \ket{0^{\otimes N}}~,
\end{equation}
where we have assumed some notion of circuit depth, $d$, such that each unitary $U_i$ contains a `layer' of entangling gates, the exact composition of which depends on the specific circuit definition.
The process for computing $c_x$ takes place in several steps;
(a) construct the tensor network corresponding the circuit;
(b) perform some purely structure dependent simplifications of the tensor network;
(c) find the contraction path for this simplified network; and
(d) actually perform the contraction using the found path.
Steps (a) and (b) are very cheap, and moreover we can re-use the path found in step (c) to contract any tensor with matching structure but different tensor entries, such as varying $x$.

\subsubsection{Gate Decompositions}\label{sec:gate-decomp}

We find that pre-processing the tensor networks with the methods from Sec.~\ref{sec:simplification} before attempting to find contraction paths is an important step, particularly for optimizers such as \texttt{QuickBB} and \texttt{Hyper-Greedy} that scale badly with the number of edges and vertices.
A tensor network for $c_x$ initially consists of: rank-1 tensors describing each of the input and output qubit states; rank-2 tensors describing single qubit gates; and rank-4 tensors describing two-qubit gates.
The first processing step is deciding how to treat the two-qubit gates.
A tensor describing such a gate can be written $g_{i_a i_b}^{o_a o_b}$, such that $i_a$ ($i_b$) is the input index and $o_a$ ($o_b$) the output index of qubit $a$ ($b$).
Whilst $g_{i_a i_b}^{o_a o_b}$ is unitary with respect to ${i_a i_b} \rightarrow {o_a o_b}$, a low rank decomposition can potentially be found by grouping the indices $\{i_a, o_a\}, \{i_b, o_b\}$ or $\{i_a, o_b\}, \{i_b, o_a\}$ and performing an SVD on the resulting matrix.
In the first case this yields two rank-3 tensors:
\begin{equation} \label{eq:gate-spatial-decomp}
    g_{i_a i_b}^{o_a o_b} =
    \sum_{\xi = 1}^{\chi}
    l^{o_a}_{i_a \xi}
    r^{o_b}_{i_b \xi} \,,
\end{equation}
\begin{center}
\includegraphics[width=0.4\linewidth]{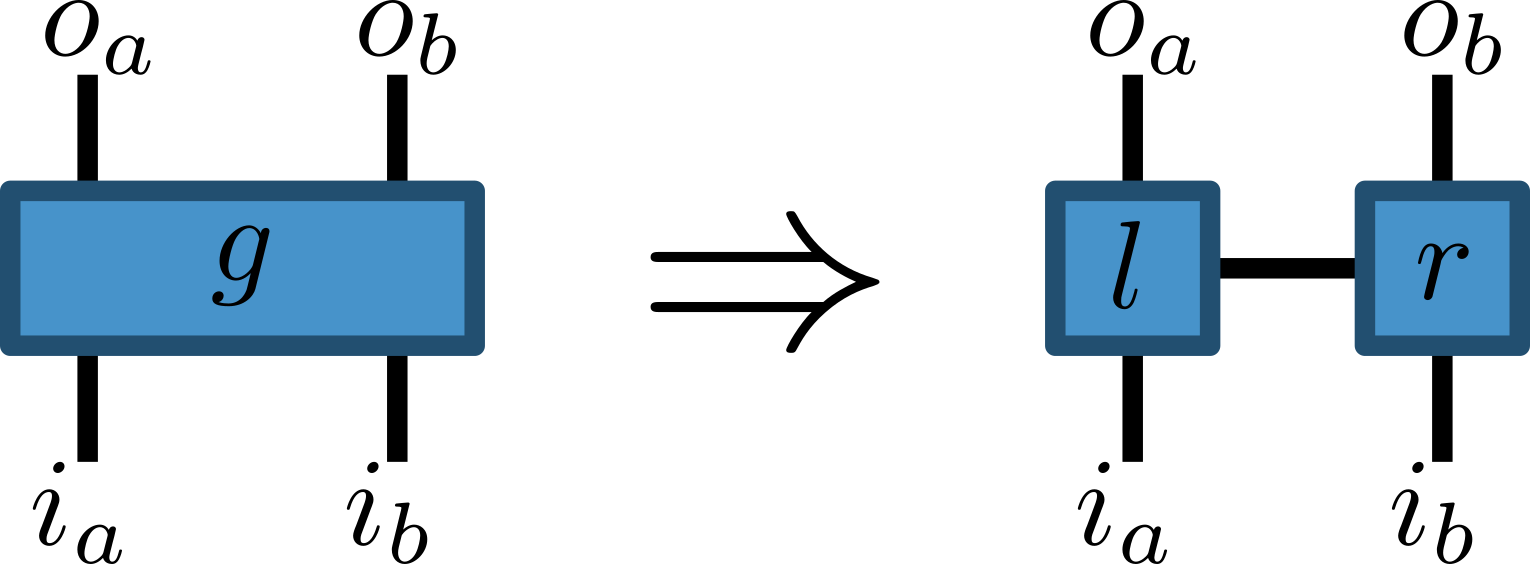}
\end{center}
where we have dropped any zero singular vectors and absorbed the remaining singular values into either of the left and right tensors $l$ and $r$, each of which is now `local' to either qubit $a$ or $b$, connected by a bond of size $\chi$.
The second case yields the same but with an effective SWAP (which can be implemented purely as a relabelling of tensor indices) of the qubit states first:
\begin{equation} \label{eq:gate-swap-decomp}
    g_{i_a i_b}^{o_a o_b} =
    \sum_{\xi = 1}^{\chi}
    \sum_{i'_a i'_b = 1}^{2}
    l^{o_a}_{i'_a \xi}
    r^{o_b}_{i'_b \xi}
    \delta^{i'_b}_{i_a} \delta^{i'_a}_{i_b}
    ~.
\end{equation}
\begin{center}
\includegraphics[width=0.4\linewidth]{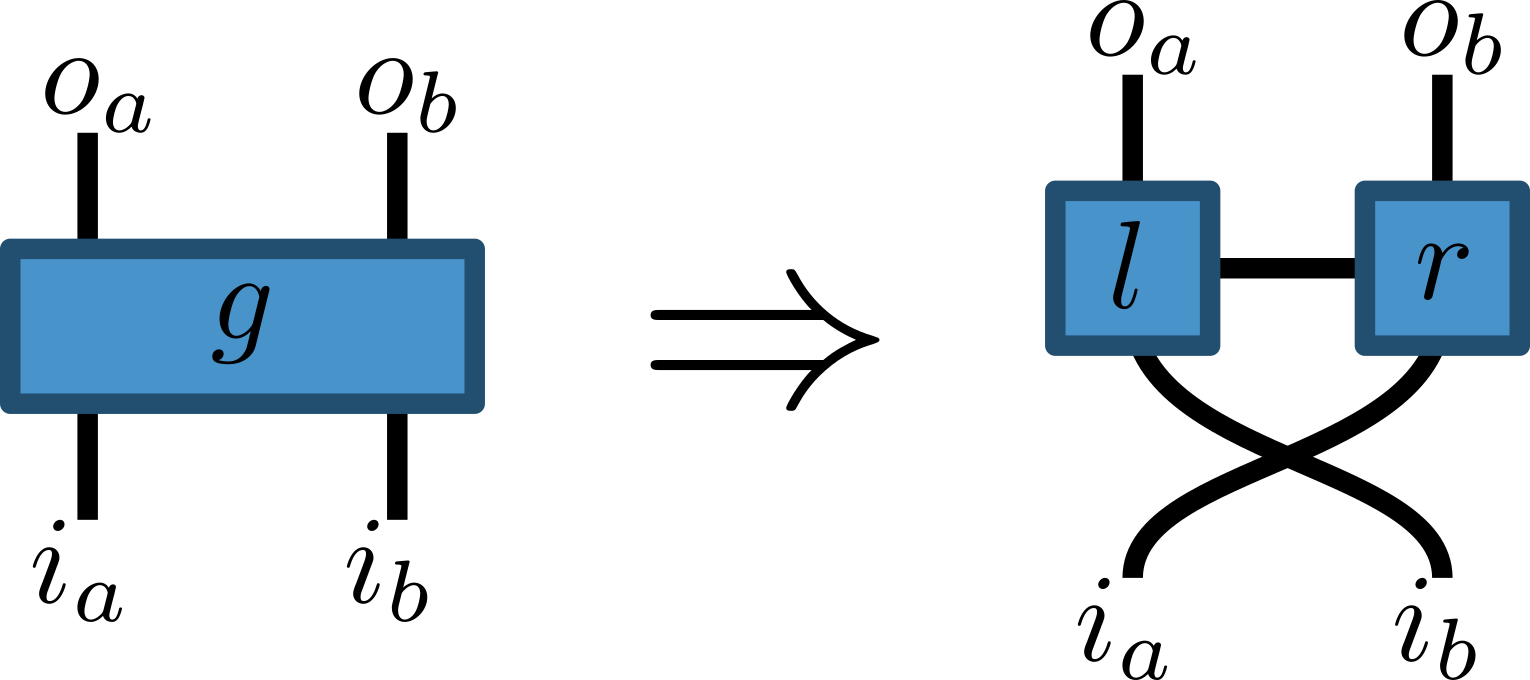}
\end{center}
The options for a gate are thus to: (a) perform no decomposition; (b) perform a \emph{spatial} decomposition -- Eq.~\eqref{eq:gate-spatial-decomp}; or (c) perform a \emph{swapped} decomposition -- Eq.~\eqref{eq:gate-swap-decomp}.
By default we only perform a decomposition if the bond dimension, $\chi$, yielded is less than 4;
all controlled gates fall into this category for a spatial decomposition, whereas the ISWAP gate for instance has $\chi=2$ for the swapped decomposition.
Such exact decompositions would also be performed automatically using the \emph{split-simplification} scheme of Sec.~\ref{sec:simplification}.
Another option is to discard small but non-zero singular values which will result in a drop in the fidelity of $c_x$~\cite{markov2018quantum,Arute2019} -- unless explicitly noted we do not perform this form of `compression'.

\subsubsection{Random Quantum Circuit Geometries}

We benchmark the contraction path optimizers against different random quantum circuits executing on three different quantum chip geometries: (i) a rectangular $7{\times}7$ lattice of 49 qubits; (ii) a 70 qubit `Bristlecone' lattice; and (iii) a 53-qubit `Sycamore' lattice.

For the first two we use the updated, harder versions of the random circuit definitions first suggested in~\cite{Boixo2018}, which are available at~\cite{sboixoGRCS}.
We adopt the notation $(1{+}d{+}1)$ for depth $d$ to emphasize that the technically first and last layer of single qubit gates (which add no real complexity) are not counted.
In both cases the entangling gate used is the controlled-$Z$ which has a $\chi=2$ spatial decomposition.

For the Sycamore architecture, we use the same circuits that were defined and also \emph{actually executed} in the recent work~\cite{Arute2019}.
Here each two-qubit gate is a separately tuned `fermionic simulation' gate which has no low-rank decomposition if treated exactly.
On the other hand, if a swapped decomposition is performed, the two smallest singular values are quite small and on average discarding them leads to a fidelity drop of a fraction of a percentage point -- for a single gate.
If this approximation is used for every single entangling gate in the circuit, however, the error is compounded.
For our main results, labelled `Sycamore-53', we thus perform no gate decomposition and consider perfect fidelity transition amplitude calculations only.
Results where the $\chi = 2$ swapped decomposition has been used we label `Sycamore-53*'.
We also note that the definition of circuit `cycles', $m$, used in~\cite{Arute2019} is about twice as hard as the rectangular and Bristlecone circuit definition of depth, $d$, since per layer almost all qubits are acted on with an entangling gate rather than approximately half respectively.

In the following table we report the number of network vertices and edges for representative depths of each circuit geometry after simplifications.
The first two columns, $|V|$, $|E|$ are for the case where hyperedge introduction is avoided, the last two columns, $\tilde{|V|}$, $\tilde{|E|}$, are for the case where the full simplification scheme introduced above has been applied. Using the ratio $\tilde{|V|}/\tilde{|E|}$ as a heuristic figure of merit, we see that the networks resulting from the Sycamore circuit model are considerably denser. One may thus anticipate that Sycamore benchmarks will be more challenging for our methods. This expectation will be borne out in Sec.~\ref{sec:qcirc-res}.

\begin{center}
\resizebox{\columnwidth}{!}{
\begin{tabular}{| c | r r | r r |}
\toprule
Circuit & $|V|$ & $|E|$ & $|\tilde{V}|$ & $|\tilde{E}|$ \\
\hline
Rectangular-$7{\times}7$ $(1 {+} 40 {+} 1)$  &~ 734  ~&~ 1101 ~&~  790 ~&~ 425 ~\\
Bristlecone-70 $(1 {+} 40 {+} 1)$   &~ 1036 ~&~ 1554 ~&~ 1086 ~&~ 574 ~\\
Sycamore-53 ($m{=}20$)          &~ 381  ~&~ 754  ~&~  381 ~&~ 754 ~\\
Sycamore-53* ($m{=}20$)         &~ 754  ~&~ 1131 ~&~ 1125 ~&~ 748 ~\\
\hline
\end{tabular}
}
\end{center}

We note that if the swap decomposition is not applied to the Sycamore circuits then no diagonal-reductions can take place and the resulting simplified tensor network is the same in both cases.

\subsubsection{2D Circuit Specific Optimizers - \texttt{qFlex/PEPs}}

Before presenting results for contraction width and cost for these random circuits, we introduce one final form of contraction path optimizer that has been successfully applied to circuits acting on 2D lattices~\cite{Villalonga2019,guo2019general}.
Here one performs the spatial decomposition of the entangling gates, regardless of rank, such that every tensor is uniquely localized above a single qubit register.
One can then contract every tensor in each of these spatial slices resulting in a planar tensor network representing $c_x$ with a single tensor per site.
Although the two works,~\cite{Villalonga2019}~and~\cite{guo2019general}, have significant differences in terms of details (and goals beyond the computation of a single perfect fidelity amplitude), the core object treated by each is ultimately this planar tensor network, which is small enough that we can report optimal contraction widths and costs for.
We call this optimizer -- which flattens the circuit tensor network into the plane before finding the optimal $W$ or $C$ \emph{from that point onwards} -- \texttt{qFlex/PEPs}.
With regards to a swapped decomposition, in order to maintain the spatial locality of the tensors this method can only benefit in the first and last layer of gates~\cite{Arute2019}.

\subsubsection{Results}\label{sec:qcirc-res}

\begin{figure*}[tb]
    \includegraphics[width=\textwidth]{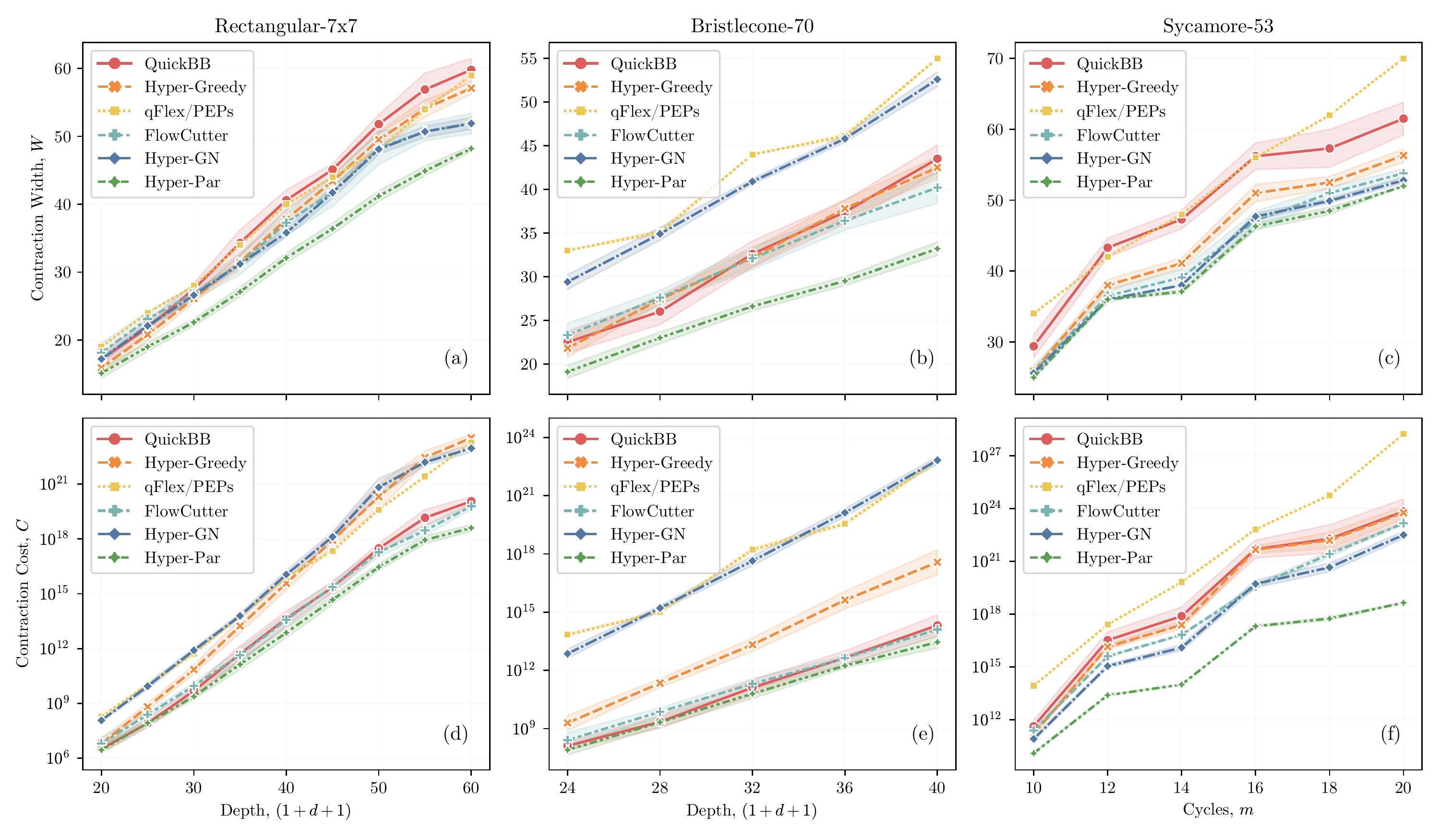}
    \caption{
        Mean contraction width (upper row) and cost (lower row) as a function of gate depth (or number of `cycles') for perfectly simulating a single output amplitude of random quantum circuits defined on three different qubit geometries -- Rectangular-7x7 (left column), Bristlecone-70 (central column) and Sycamore-53 (right column) -- for different contraction path optimizers each allowed an hour to search.
        The shaded regions show the standard deviation across 10 random circuit instances, non-zero despite the network structure of each being identical, since all optimizers but the \texttt{qFlex/PEPs} approach are naturally stochastic.
    }
    \label{fig:rand-circs}
\end{figure*}

In Fig.~\ref{fig:rand-circs}(a)-(f) we report the mean contraction width, $W$, and cost, $C$, for each geometry and optimizer as a function of circuit depth, $d$, or cycles, $m$.
For these large tensor networks we allow each optimizer one hour to search for a contraction path.
While this is not an insignificant amount of time, we note that many optimizers converge to their best contraction paths much quicker, and moreover that contraction paths can be re-used if only changing tensor values from run to run.
We show the variance in $W$ and $C$ across 10 instances, despite the fact the tensor network structure is the same, since all the optimizers aside from \texttt{qFlex/PEPs} are naturally stochastic.

We first note that across the board, the \texttt{Hyper-Par} optimizer again performs best, with little variance from instance to instance.
Performance of the remaining optimizers is more difficult to rank.
The tensor network simplification scheme employed here results in significant improvement over previous results even when using \texttt{QuickBB} to perform the actual path optimization, particularly when $|E|$ or $|\tilde{E}|$ is moderate.
As the tensor networks get larger \texttt{QuickBB} is consistently outperformed by the other line-graph based optimizer \texttt{FlowCutter}.

For the Rectangular-7x7 and Bristlecone-70 circuits, which both use a CZ entangling gate, the diagonal reduction of tensors greatly simplifies the tensor networks.
The methods that make use of this, aside from \texttt{Hyper-Greedy}, perform best here, with similar values of $C$, though interestingly \texttt{Hyper-Par} is able to target a lower contraction width.
\texttt{Hyper-GN} and \texttt{qFlex/PEPs} do not use the diagonal simplification and here show similar performance.

In the case of Sycamore-53 the entangling fSim~\cite{Kivlichan2018} gates are close to but not exactly ISWAP gates.
As a result there are no diagonal reductions to be made and the simplified tensor network has no hyper-edges.
Whilst \texttt{FlowCutter}, \texttt{Hyper-GN} and \texttt{Hyper-Par} find similar contraction widths, \texttt{Hyper-Par} achieves a much lower contraction cost.
This is likely due to its ability to search imbalanced partition contraction trees such as `Schr\"{o}dinger style' (full wavefunction) evolution.
Note that for the entangling gates an approximate swapped $\chi{=}2$ decomposition can be made, resulting in a drop in fidelity based on how many of the $m$ layers of gates this is applied to.
The \texttt{qFlex/PEPs} method results in~\cite{Arute2019} make use of this in the first and last layer of gates for a drop in total fidelity of ${\sim} 5\%$ that reduces $W$ by ${\sim} 4$ and $C$ by ${\sim} 2^4$.
We only show the exact results here so as to compare all methods on exactly the same footing.
If the swapped decomposition is used for \emph{all} layers (Sycamore-53*) then at $m{=}20$ the corresponding drop in total fidelity is likely to be ${\sim}50\%$.
For the best performing optimizers in Fig.~\ref{fig:rand-circs}(c) and (f) we find little gain in doing so.
We also emphasize that for the highest values of $m$, the estimates for classical computation cost in~\cite{Arute2019} are not based on the \texttt{qFlex}~\cite{Villalonga2019} simulator and moreover involve the unbiased \emph{sampling} of many bit-strings at low fidelity.

\subsection{Practical Performance}

In this final results section, we examine how the high quality contraction paths obtained so far transform into practical performance.
Whilst the contraction cost estimates the time complexity of contracting a tensor network, this is irrelevant if the contraction width is too large to fit the computation into available memory.
One method to bring down the space requirement of any contraction is \emph{slicing}, also known as `variable projection'~\cite{Chen2018} or `bond cutting'~\cite{Villalonga2019}.

\subsubsection{Slicing}

\begin{figure}[tb]
    \centering
    \includegraphics[width=\linewidth]{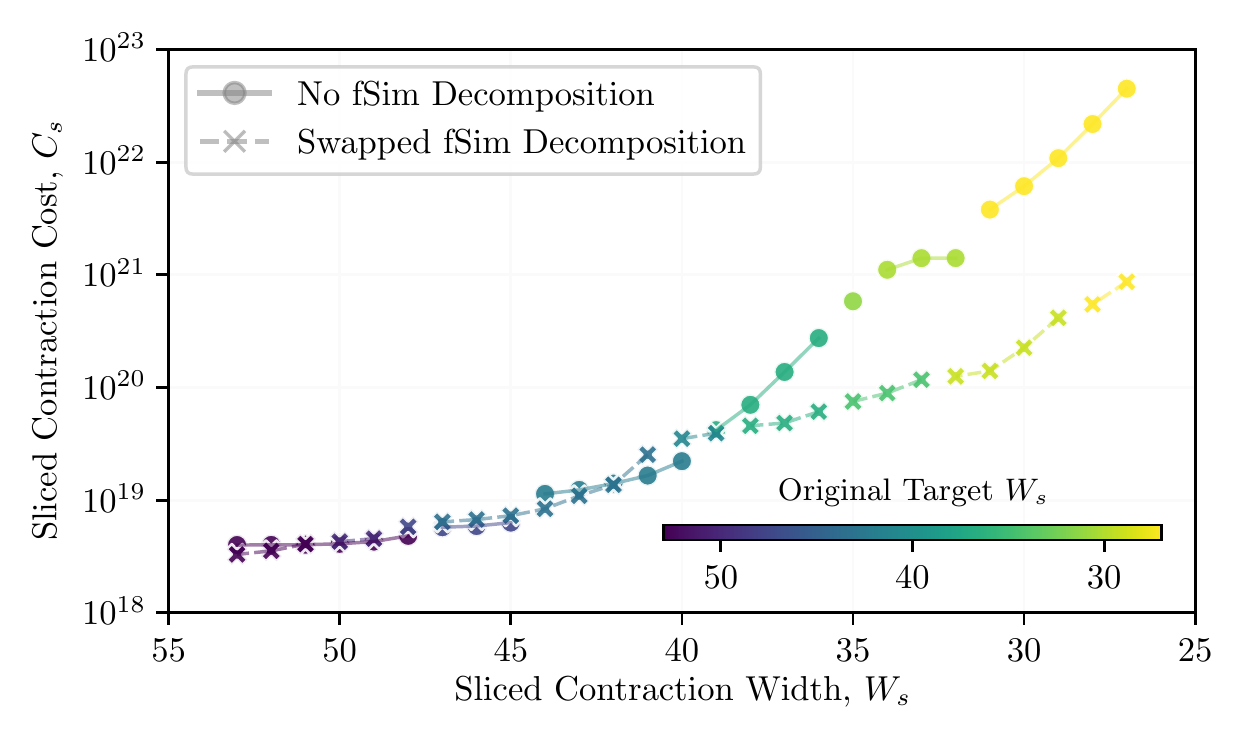}
    \caption{Sliced contraction width vs. cost for computing a single amplitude of an $m=20$ cycle random Sycamore circuit using the \texttt{Hyper-Par} optimizer.
    The two sets of markers correspond to how the fSim gate is applied: either the tensor is not decomposed and left rank-4, or the two qubits are first swapped and then a decomposition is performed with only the two dominant singular values kept -- Sycamore-53 and Sycamore-53* respectively.
    Points joined by a line are sliced from the same original contraction tree, with colour given by what contraction width was being targeted by the Bayesian optimization.
    Each optimizer targeting a particular $W_s$ was allowed 1 hour to search.
    }
    \label{fig:sycm-sliced}
\end{figure}

A tensor network can always be thought of as $|E|$ nested summations of the product of the entries of the $|V|$ tensors.
Such an expression is associative and a contraction tree is equivalent to a re-arrangement of the summations and the insertion of a sequence of $|V|-1$ parentheses defining intermediate tensors to form.
However, we can also choose to perform any subset of the summations \emph{last}, moving them back to the exterior of the expression.
We'll call the corresponding set of indices $s_\mathrm{sliced}$.
For each fixed value of this exterior sum, the remaining expression corresponds to a tensor network of $|V|$ nodes, but with all the edges in $s_\mathrm{sliced}$ removed.
In each network, the fixed value of indices corresponds to taking \emph{slices} of any tensors with those indices.
The total number of such sliced tensor networks is then $d_\mathrm{sliced} = \prod_{e \in s_\mathrm{sliced}} w(e)$, each of which can be contracted independently, optionally using the same tree as the original network.

The advantage of doing this is twofold: (i) the contraction width and thus required memory of each sliced tensor network, $W_s$, is generally reduced; and (ii) the sum over independent contractions is `embarrassingly parallel' and so can be easily distributed.
The disadvantage is that the contraction cost of each sliced tensor network generally increases beyond $C / d_\mathrm{sliced}$ (due to redundantly repeated contractions) meaning the \emph{total} sliced cost, $C_s$, rises.
Choosing which indices to slice is thus a balancing act between reducing the memory footprint without increasing the cost too much.

We employ a method similar to~\cite{Chen2018} to choose which indices to slice.
Given a contraction tree $B$, it is simple to compute the new width and cost with any index removed using Eqs.~\eqref{eq:edgecon}~and~\eqref{eq:vertexcon}.
We greedily choose single indices to slice based on this, repeating the process until the sliced contraction tree width reaches the desired target.
Repeating this process a few times with a slight randomization to the cost score allows us to sample a moderate number of combinations for $s_\mathrm{sliced}$ and choose whichever achieves target $W_s$ whilst minimizing $C_s$.
Crucially, we can slice trial contraction trees and report $C_s$ \emph{within} the Bayesian optimization loop, thus explicitly targeting paths which slice well.

In Fig.~\ref{fig:sycm-sliced} we demonstrate the effect of different levels of slicing for the deepest Sycamore-53 circuit ($m{=}20$), with either no fSim gate decomposition, or the approximate $\chi{=}2$ gate decomposition for all layers (Sycamore-53*), which now shows an appreciable benefit.
We allow the optimizer an hour to find paths with the lowest $C_s$ for a given target $W_s$.
If a path targeting a neighbouring $W_s$ achieves a lower $C_s$, this is shown instead, and the points connected by a line.
One can see that the required memory can be brought down by a factor of $\sim16,000$ whilst keeping the FLOPs increase $<10$.
Across this same range performing the swapped decomposition yields no benefit.
Beyond that, the increase to $C_s$ becomes significant, with the swapped decomposition becoming advantageous for heavily sliced contractions.
For reference, $W_S \sim 27$ is required to fit a contraction on a standard consumer GPU.
Interestingly, the paths which achieve lowest overall $C_s$ when targeting a large $W_s$ (dark purple), are not good candidates for heavy slicing (yellow).
Instead, the Bayesian optimizer targets a variety of different paths specific to each level of slicing.

\subsubsection{Benchmarks}

\begin{table*}[t]
\begin{center}
\resizebox{\textwidth}{!}{
\begin{tabular}{| c | c | c | c | c | c |}
 \toprule
 Circuit &
 time (sec) &
 $C_s$ &
 Slicing Overhead ($C_s / C_\mathrm{best}$) &
 $d_\mathrm{sliced}$ &
 FLOPs Efficiency \\
 \hline
 Bristlecone-70 $(1 {+} 32 {+} 1)$
 ~&~ $4.18 \times 10^{-1}$ ~&~ $4.91 \times 10^{10}$ ~&~ $1.24 \times$ ~&~ $2$ ~&~ 31.0\% ~\\
 Bristlecone-70 $(1 {+} 36 {+} 1)$
 ~&~ $1.74 \times 10^{1}$ ~&~ $2.06 \times 10^{12}$ ~&~ $1.05 \times$ ~&~ $2^{5}$ ~&~ 31.1\% ~\\
 Bristlecone-70 $(1 {+} 40 {+} 1)$
 ~&~ $2.77 \times 10^{2}$ ~&~ $3.14 \times 10^{13}$ ~&~ $1.65 \times$ ~&~ $2^{8}$ ~&~ 29.9\% ~\\
 \hline
 Rectangular-$7{\times}7$ $(1 {+} 32 {+} 1)$
 ~&~ $3.38 \times 10^{-1}$ ~&~ $2.84 \times 10^{10}$ ~&~ $1.49 \times$ ~&~ $2$ ~&~ 22.2\% ~\\
 Rectangular-$7{\times}7$ $(1 {+} 40 {+} 1)$
 ~&~ $4.80 \times 10^{1}$ ~&~ $8.12 \times 10^{12}$ ~&~ $1.35 \times$ ~&~ $2^{7}$ ~&~ 44.6\% ~\\
 Rectangular-$7{\times}7$ $(1 {+} 48 {+} 1)$
 ~&~ $~^*9.40 \times 10^{4}$ ~&~ $1.20 \times 10^{16}$ ~&~ $1.33 \times$ ~&~ $2^{18}$ ~&~ 33.7\% ~\\
 \hline
 Sycamore-53 ($m{=}12$)
 ~&~ $5.74 \times 10^{2}$ ~&~ $1.80 \times 10^{14}$ ~&~ $7.51 \times$ ~&~ $2^{9}$ ~&~ 82.6\% ~\\
 Sycamore-53 ($m{=}14$)
 ~&~ $~^*4.98 \times 10^{3}$ ~&~ $1.37 \times 10^{15}$ ~&~ $13.6 \times$ ~&~ $2^{12}$ ~&~ 72.8\% ~\\
 Sycamore-53 ($m{=}16$)
 ~&~ $~^*8.01 \times 10^{6}$ ~&~ $2.41 \times 10^{18}$ ~&~ $13.0 \times$ ~&~ $2^{22}$ ~&~ 79.4\% ~\\
 Sycamore-53 ($m{=}18$)
 ~&~ $~^*8.18 \times 10^{7}$ ~&~ $2.64 \times 10^{19}$ ~&~ $\boldsymbol{42.6 \times}$ ~&~ $2^{24}$ ~&~ 85.2\% ~\\
 Sycamore-53 ($m{=}20$)
 ~&~ $~^*9.74 \times 10^{10}$ ~&~ $3.10 \times 10^{22}$ ~&~ $\boldsymbol{6410 \times}$ ~&~ $2^{34}$ ~&~ 84.1\% ~\\
 \hline
 Sycamore-53* ($m{=}12$)
 ~&~ $7.87 \times 10^{2}$ ~&~ $2.42 \times 10^{13}$ ~&~ $1.67 \times$ ~&~ $2^{9}$ ~&~ \textbf{8.16\%} ~\\
 Sycamore-53* ($m{=}14$)
 ~&~ $~^*2.92 \times 10^{3}$ ~&~ $2.53 \times 10^{14}$ ~&~ $2.63 \times$ ~&~ $2^{12}$ ~&~ 22.9\% ~\\
 Sycamore-53* ($m{=}16$)
 ~&~ $~^*3.01 \times 10^{6}$ ~&~ $3.43 \times 10^{17}$ ~&~ $7.43 \times$ ~&~ $2^{22}$ ~&~ 30.1\% ~\\
 Sycamore-53* ($m{=}18$)
 ~&~ $~^*2.66 \times 10^{7}$ ~&~ $3.62 \times 10^{18}$ ~&~ $11.3 \times$ ~&~ $2^{24}$ ~&~ 36.0\% ~\\
 Sycamore-53* ($m{=}20$)
 ~&~ $~^*7.17 \times 10^{9}$ ~&~ $1.50 \times 10^{21}$ ~&~ $\boldsymbol{431 \times}$ ~&~ $2^{32}$ ~&~ 55.3\% ~\\
 \hline
\end{tabular}
}
\end{center}
\caption{\label{tab:benchmark} Benchmark times and other information for computing a single amplitude of random circuits, in single precision.
The time shown is for the contraction only, using a NVIDIA Quadro P2000 for which a target sliced contraction width $W_s = 27$ suffices for its 5GB of memory.
Times with an asterisk are estimates extrapolated from computing the first $100 $ of $d_\mathrm{sliced}$ contractions.
The sliced cost, $C_s$, is always higher than the best cost without slicing, $C_\mathrm{best}$ (shown in Figs.~\ref{fig:rand-circs}(d), (e) and (f)).
As such, the `slicing overhead' indicates the inefficiency induced by squeezing the contraction into 5GB.
The FLOPs efficiency compares the theoretical single precision performance of the Quadro P2000, 3.031~teraFLOPs, with $(8 C_s / \mathrm{time})$.
}
\end{table*}

To demonstrate that the contraction paths and calculated costs translate well into real world performance, we here report actual times for contracting a single perfect fidelity amplitude on a single GPU for various circuits.
All tensor network manipulations and contractions were performed using \texttt{quimb}~\cite{gray2018quimb}.
For each run, we allow the path optimizer to search for 1 hour in the space of paths sliced to $W_s = 27$.
We then compile the resulting contraction using \texttt{JAX}~\cite{jax2018github} and run it on a NVIDIA Quadro P2000 which has 5GB of memory and theoretical single precision performance of 3.031~teraFLOPs.
Both the path finding and compilation time are one-off costs per circuit and the times we report are only for performing the contraction.
All the examples shown require some degree of slicing to fit onto the GPU, so we also show the sliced cost and how this compares to the best non-sliced cost.
This \emph{slicing overhead} is the increase in cost induced by squeezing the contraction into 5GB of memory.
Finally we compare the achieved FLOP rate to theoretical maximum for the GPU.

The results are shown in Tab.~\ref{tab:benchmark}. For this specific task, and to the best of our knowledge, these generally represent state-of-the-art performance.
For the rectangular and Bristlecone geometries, there is little inefficiency induced by slicing the contractions down to fit into memory.
On the other hand, the performance extracted from the GPU via \texttt{JAX} is not great, likely due to the fact that the corresponding tensor networks have hyper-edges resulting in pairwise contractions that do not dispatch to matrix-matrix multiplication.
For Sycamore-53, there are no hyper-edges and the realised FLOP rate is close to the theoretical limit of the GPU.
On the other hand, there is much greater inefficiency induced by slicing the contractions down to $W_s=27$.
For $m=20$ this overhead is very significant, representing the far right point of Fig.~\ref{fig:sycm-sliced}.
From that same figure it can be seen that performing the swapped decomposition alleviates the slicing overhead, and indeed we find this to be the case with the Sycamore-53* benchmarks, though the introduction of hyperedges again lowers the FLOPs efficiency.
From Fig.~\ref{fig:sycm-sliced} it can also be seen that there are steady gains to be made by allowing a higher $W_s$, either through simply more memory or moving to a distributed computing setting.
In the latter case, sliced indices might instead be suggestive of how to partition the initial tensors.

\subsubsection{Estimated Runtime of Sampling Sycamore}

So far the reported contraction costs have been those associated with computing a single transition amplitude of the circuit with perfect fidelity. In order to classically simulate taking $M$ approximately unbiased samples of the circuit at fidelity $f$ we require a few extra steps.

Firstly, since the circuits in \cite{Arute2019} are chaotic, rather than computing the probability distribution over all bit strings we can assume that any marginal distribution on $N_u$ qubits is uniform once a certain number of qubits, $N_f$, are traced out.
We can therefore uniformly select a bitstring, $x_u$, to fix the final state of the first $N_u$ qubits to, and simply compute the the probability distribution over the remaining $N_f$ qubits conditioned on $x_u$. If we sample $x_f$ from this final marginal, with probability given by
\begin{equation}
p(x_f | x_u) = \frac{1}{Z} |\left(\langle x_u | \otimes \langle x_f|\right) \left| \psi \right\rangle |^2~~,
\end{equation}
for some normalization $Z$, then the bitstring $x_u x_f$ will be an approximately unbiased sample from $\psi$. Crucially, if $N_f$ is small enough and the qubits chosen sensibly, the cost of computing $p(x_f | x_u) $ for all $2^{N_f}$ final bitstrings, $\{x_f\}$, is generally very similar to that of computing a single amplitude.
The TN contracted here is now like Eq.~\eqref{eq:trans-amp} but with $N_f$ output indices left open.
Secondly, we can simulate fidelity $f$ by sampling from $\rho = (1 - f) \mathbf{1}/{2^{N}} + f \ket{\psi}\bra{\psi}
$, which in practice means yielding with probability $f$ a bitstring from $\psi$ but uniformly sampled bitstrings the remainder of the time.
If $\psi$ itself has fidelity $g$ we can take $f \rightarrow \tilde{f} = f/g$ to compensate.

In \cite{villalonga2019establishing}, sliced tensor network contractions were performed on the supercomputer Summit, which we can use as a basic reference to estimate the cost of classically simulating the supremacy task~\cite{Arute2019}.
The contractions here involve no caching or communication between nodes, but do make use of out-of-core contractions, enabling a sliced width of $W_s{=}32$.
If we take $N_f=6$ on Sycamore-53* (qubits 10, 17, 26, 36, 27, 18) we find a sliced contraction cost of $C_s=10^{20.17}$, which in reference to Fig.~\ref{fig:sycm-sliced} is indeed very similar to single amplitude cost.
To sample $M=1,000,000$ bitstrings at fidelity $f=0.002$ with wavefunction fidelity $g \sim 0.5$ (due to swapped fSim decompositions) we thus need to perform $\frac{Mf}{g} = 4000$ contractions.
In \cite{villalonga2019establishing} a sustained rate of 281 petaFLOPs was achieved, corresponding to a $68\%$ FLOPs efficiency.
Taking this as an upper bound we get
\begin{equation}
    \frac{8 \times 10^{20.17} \times 4000}{281\times 10^{15}} = 195~\mathrm{days}
\end{equation}
to perform the task, or $241~\mathrm{days}$ if we take the FLOPs efficiency of $55\%$ from Table~\ref{tab:benchmark}.
Both represent a speed-up of over $10,000\times$ with regard to the estimated time of performing the sampling task classically in \cite{Arute2019}.
Reducing the slicing overhead via distributed contraction or better slicing algorithms might well bring this down further.
Another interesting prospect is whether fidelity $f$ can be targeted via an algorithm with better speed-up than the basic $1/f$ here.

\section{Summary and conclusion}\label{sec:conclusion}

We have introduced heuristic algorithms for the contraction of arbitrary tensor networks that show very good performance across a range of benchmarks.
These explicitly construct a contraction tree and target the cost of all operations in the contraction.
Through a stochastic hyper-optimization over the parameters of each of the algorithms, we obtain near-optimal contraction paths that yield exponential speedups over the state-of-the-art contraction algorithms. We find that the contractor based on hypergraph partitioning, in particular, often outperforms all other methods. We demonstrated how this translates to superior performance in the simulation of computing amplitudes on Google quantum chips. In particular, we have estimated a speed-up of over 10,000$\times$ compared to the original expectation for the classical simulation of the Sycamore `supremacy' circuits.

While our contraction path optimization methods find near-optimal paths for all the benchmarks we have tried, in some cases their advantage over less sophisticated methods is modest. The reason is that for the associated families of graphs, e.g., planar graphs and grids, good contraction paths are easy to find by either inspection or naive greedy search. In all other cases, however, the performance advantage over already established tensor network contraction methods is much more pronounced.

Due to the generality of tensor networks, our results can help advance applications in a variety of fields. The algorithms introduced here can be directly employed in the calibration of ever larger quantum chips, with techniques such as cross-entropy benchmarking. They can also form the basis of decoders based on contraction of disordered tensor networks, an increasingly important component of quantum error correction~\cite{Bravyi2014,Ferris2014,Chubb2018}. They are also immediately applicable to computational tasks related to artificial intelligence, such as inference and model counting~\cite{Dudek2019}, to reliability engineering~\cite{Duenas-Osorio2018}, and more. While without slicing our hyper-optimized contractors essentially achieve optimality in practice for the vast majority of problem instances, improvements in slicing strategies may be attainable. Finally, incorporating controllable schemes for approximate contractions into the methodology introduced here is a promising domain of future research. At a most basic level, one can easily perform truncated singular value decompositions after every few contraction steps of our algorithms, which would further increase performance (even to polynomial time and space, depending on the truncation scheme) at the expense of accuracy. This may enable computations of observables in quantum many-body systems with disorder or irregular geometries, which have so far remained mostly out of the reach of tensor network methods.

\begin{acknowledgments}
We thank S.~Boixo and B.~Villalonga for useful feedback on the manuscript. JG acknowledges the Samsung Advanced Institute of Technology Global Research Partnership. SK was supported in part by funding from the Canada First Research Excellence Fund.
\end{acknowledgments}

\bibliography{auto}
\bibliographystyle{apsrev4-1}

\end{document}